\documentclass[useAMS,usenatbib]{mn2e}
\usepackage{epsfig,graphicx,lscape}
\title[Interstellar Na\,{\sc i} UV, Ti\,{\sc ii} and Ca\,{\sc ii}]{Early-type stars observed in the ESO UVES Paranal Observatory Project: I -- Interstellar
Na\,{\sc i} UV, Ti\,{\sc ii} and Ca\,{\sc ii} K observations.} 

\author[I. Hunter et al.]
         {I. Hunter$^{1,2}$\thanks{email: i.hunter@qub.ac.uk. \hspace*{0.3cm} 
          \newline Based on 
          observations taken at UT2, Kueyen, Cerro Paranal, Chile, 
           ESO DDT programme 2665.D-5655, UVES Paranal Observatory Project.}
          J.~V. Smoker$^{1,2}$, F.~P. Keenan$^{1}$, C. Ledoux$^{2}$, 
          E. Jehin$^{2}$, \newauthor
          R. Cabanac$^{3}$, C. Melo$^{2}$, S. Bagnulo$^{2}$.
          \\
          \\
       $^{1}$Astrophysics and Planetary Science Research Division,
            Department of Physics and Astronomy,
            The Queen's University of Belfast, \\
            University Road, Belfast, BT7 1NN,
            U.K. \\
       $^{2}$European Southern Observatory,
            Alonso de Cordova 3107,
            Casilla 19001,
            Vitacura,
            Santiago 19, Chile \\
       $^{3}$Canada-France-Hawaii Telescope Corporation,
        65-1238 Mamalahoa Hwy.
        Kamuela, Hawaii 96743, USA
                 \\
}

\date{Accepted
      Received
      in original form }

\def\LaTeX{L\kern-.36em\raise.3ex\hbox{a}\kern-.15em
    T\kern-.1667em\lower.7ex\hbox{E}\kern-.125emX}

\pubyear{}

\begin{document}
\label{firstpage}
\maketitle
\begin{abstract}

We present an analysis of interstellar Na\,{\sc i} 
($\lambda_{\rm air}$=3302.37\AA, 3302.98\AA), 
Ti\,{\sc ii} ($\lambda_{\rm air}$=3383.76\AA) and Ca\,{\sc ii} K 
($\lambda_{\rm air}$=3933.66\AA) absorption features  
for 74 sightlines towards O- and B-type stars 
in the Galactic disc. The
data were obtained from the UVES Paranal Observatory Project, at a 
spectral resolution of 3.75~km~s$^{-1}$ and 
with mean signal to noise ratios per pixel of
260, 300 and 430 for the 
Na\,{\sc i}, Ti\,{\sc ii} and Ca\,{\sc ii} observations, 
respectively. Interstellar features 
were detected in all but one of the Ti\,{\sc ii} sightlines and all of the 
Ca\,{\sc ii} sightlines. The
dependence of the column density of these three species with distance, height
relative to the Galactic plane, H\,{\sc i} column density, reddening and
depletion relative to the solar abundance has been investigated. We also examine
the accuracy of using the Na\,{\sc i} column density as an indicator 
of that for H\,{\sc i}. 
In general we find similar strong correlations for both
Ti and Ca, and weaker correlations for Na. Our results confirm the general belief
that Ti and Ca occur in the same regions of the interstellar medium
and also that the Ti\,{\sc ii}/Ca\,{\sc ii} ratio is constant over all
parameters. We hence conclude that the absorption properties of Ti
and Ca are essentially constant under the general interstellar medium
conditions of the Galactic disc.

\end{abstract}


\section{Introduction}                                           \label{s_intro}

The study of the interstellar medium in our Galaxy has a long history, 
dating back a century when interstellar Ca\,{\sc ii} absorption was
first detected (Hartmann 1904). Historically, ground-based work has 
concentrated on the Ca\,{\sc ii} K and Na\,{\sc i} D lines at around
3933~\AA\ and 5980~\AA. Studies 
such as Albert et al. (1993), Sembach, Danks \& Savage (1993), Welty, Hobbs 
\& Kulkarni (1994), Welty, 
Morton \& Hobbs (1996) and Smoker et al. (2003) have observed the 
Ca\,{\sc ii} K and Na\,{\sc i} D lines in an attempt to provide information 
on the structure and physical characteristics of interstellar clouds, in 
particular their spatial distribution (e.g. Lallement et al. 2003), kinematics 
(e.g. Sembach \& Danks 1994) and depletion patterns of these elements
(e.g. Crinklaw, Federman 
\& Joseph 1994; Wakker \& Mathis 2000). The general 
situation can be summarised as follows; Ca\,{\sc ii}, with its ionisation 
potential of $\sim$ 11.9 eV, is typically a trace ionisation stage in diffuse, 
neutral clouds, plus a more extended intercloud medium. However Na\,{\sc i} 
is found to trace the neutral clouds only. Ca\,{\sc ii} is thought to be 
heavily depleted onto dust in denser clouds, with Na\,{\sc i} a little-depleted
element (Welty et al. 1996).

Other interstellar lines which have been studied much
less frequently are 
the UV Na\,{\sc i} doublet at 
$\lambda_{\rm air}$=3302.37\AA, 3302.98\AA,
and Ti\,{\sc ii} at 
$\lambda_{\rm air}$=3383.76\AA. The relative paucity of observations is 
for two main reasons. First, the oscillator strengths of 
$f$=0.0092, 0.0046 and 0.358 for the Na\,{\sc i} doublet and Ti\,{\sc ii}, respectively, 
are somewhat smaller than for the Ca\,{\sc ii} K ($f$=0.627) and 
Na\,{\sc i} D lines ($f$=0.318, 0.631). Second, both the Na\,{\sc i} UV doublet and 
Ti\,{\sc ii} lines lie toward the blue end of the optical range, where 
detectors are less efficient. Although the weakness of the lines clearly 
makes them more difficult to observe, it has the advantage that
(especially in the case of Na\,{\sc i} UV doublet) saturation effects are 
much less of an issue. Hence these transitions
can be used to probe quite dense 
regions of the ISM, without having to apply large corrections for the 
effects of line saturation. In the case of Ti\,{\sc ii}, the ionisation 
potential of 13.57 eV is nearly the same as that of H\,{\sc i}, so that
this species is the dominant ionisation stage in H\,{\sc i} regions and no 
ionisation corrections are necessary to obtain the total abundance of 
Ti in these regions. This is in contrast to Ca\,{\sc ii} K in warm 
neutral or weakly ionized gas, where the species
is not the dominant ionisation stage (e.g. Sembach et al. 2000).

Previous observations of the Na\,{\sc i} UV lines have been published by only 
a handful of authors including Hobbs (1978), Vallerga et al. (1993) 
and Welsh et al. (1997); corresponding results for Ti\,{\sc ii} appear in
Stokes (1978), Hobbs (1984) and Welsh et al. (1997). In total approximately 
110 Ti\,{\sc ii} sightlines have been studied to date, although many of these
were non-detections. 

The results for the Na\,{\sc i} UV lines are complementary to previous studies 
of Na\,{\sc i} D. Data for Ti\,{\sc ii} thus far indicate that this element appears 
to be associated with a smoothly distributed neutral medium. Although,  
unlike Ca\,{\sc ii}, Ti\,{\sc ii} is the dominant ionisation stage in neutral gas, 
the Ti\,{\sc ii} abundance shows a similar amount of scatter when plotted 
against $N_{\rm HI}$ (Wakker \& Mathis 2000), with the 
Ti\,{\sc ii}/Ca\,{\sc ii} abundance ratio being constant over a range of 
interstellar density conditions (Welsh et al. 1997). 

The current paper concerns observations of the Na\,{\sc i} UV doublet, Ti\,{\sc ii} and 
Ca\,{\sc ii} K towards a sample of 74 
early-type stars in the Southern Hemisphere. The wavelengths and transition 
strengths for the studied lines are shown in Table~\ref{t_wavef}.

\begin{table}
\begin{center}
\caption[]{Main transitions studied.}
\label{t_wavef}
\begin{tabular}{lcr} \hline
Transition & $\lambda_{\rm air}$ & $f$-value \\
           &  (\AA)              &           \\
\hline
Na\,{\sc i}    & 3302.368  & 0.00921         \\
Na\,{\sc i}    & 3302.978  & 0.00460         \\
Ti\,{\sc i}    & 3383.759  & 0.358           \\
Ca\.{\sc ii}   & 3933.661  & 0.627           \\
\hline
\end{tabular}
\end{center}
\end{table}

In contrast to previous surveys which generally have signal-to-noise (S/N)
ratios in the Na\,{\sc i}
UV lines and Ti\,{\sc ii} of around 40, the current paper utilises data that 
typically have S/N ratios exceeding 250. This enables features with 
equivalent widths of $<$ 1~m\AA\ to be detected. This contrasts with 
the only previously existing Southern Hemisphere survey of Ti\,{\sc ii} by 
Welsh et al. (1997; hereafter W97) in which, with a S/N of 40, 18 out of 42 
of the observed sightlines in Ti\,{\sc ii} were non-detections. 

In Sect. 2 we describe the sample of stars observed, while
Sect. 3 discusses 
the data analysis and methods used to obtain equivalent widths, $b$-values, 
velocities of components and column densities, including an estimate of 
the errors in the derived parameters and comparison with previous work. 
Sect. 4 presents correlations between the derived column densities and 
parameters such as reddening and height above the Galactic disc ($z$). 
Finally, in Sect. 5 we 
list the main conclusions. A future paper will deal in more 
detail with the search for intermediate and high velocity gas along the 
current sightlines. 

\section{The sample}                                              \label{s_samp}

We have extracted spectra for three interstellar species, namely the 
Na\,{\sc i} UV doublet, Ti\,{\sc ii} and Ca\,{\sc ii} K 
along the sightlines to 74 O- and B-type field stars from 
$An$ $Atlas$ $of$ $High$-$Resolution$ 
$Spectra$ $of$ $Stars$ $across$ $the$ $Hertzsprung$-$Russell$ $Diagram$,
available from the UVES  Paranal Observatory Project (POP; Bagnulo et al. 2003).
Although Ca\,H is also present in the spectra, it is on the wing of an 
H\,{\sc i} line, so was not used in the bulk of the analysis. 

High resolution spectra of the stars in the POP survey 
were obtained using the Ultraviolet and
Visual Echelle Spectrograph (UVES) mounted on the 8.2-m UT2 (Kueyen) telescope at
the European Southern Observatory (ESO), Paranal, Chile. The observations were
taken between February 2001 and February 2003, with the field stars generally
being observed during twilight. Further details of the observations and data
reduction methods are available at http://www.sc.eso.org/santiago/uvespop/.
The combination of bright stars, an efficient detector and an 8.2-m telescope results 
in very high S/N
ratios for the merged spectra, 
the average per pixel being 260, 300 and 430 in
the regions of the Na\,{\sc i} UV, Ti\,{\sc ii} and Ca\,{\sc ii} K lines,
respectively. There are 2 pixels per resolution element.
For all the spectra, the 
velocity resolution is 3.75\,km~s$^{-1}$. 

Spectra were selected on the criteria
that the continuum in the region of interest was relatively unaffected by
stellar features. As the stars of our sample are O- and B-type, in most cases
the stellar lines were sufficiently broad that they could easily be
classified as stellar and not interstellar lines. In Table~\ref{t_stars} we 
list
the field stars
in our sample, and also their spectral types,
peculiarities and visual magnitudes as listed by UVES POP. Spectral
types range from O4\,V to B9.5\,V, and all objects are brighter than 
$m_{v}$ = 8.0. 
Note that the peculiarities are taken from UVES POP and hence are not listed
in the standard international nomenclature; http://sc.eso.org/santiago/uvespop/
should be consulted for full details. 

In Table~\ref{t_stars} we also include the
reddening, $E(B-V)$, along the line-of-sight and the
spectroscopic distance estimate to each star. 
The spectra obtained from UVES POP had already been shifted to the
helio-centric frame of rest. Each spectrum was furthermore 
shifted to the kinematic Local
Standard of Rest using the corrections generated by the program {\sc rv} 
(Wallace \& Clayton 1996). In the majority of cases the LSR velocity of the stars were
calculated from the positions of strong stellar lines such as Si\,{\sc ii}
3852~\AA\ or
Si\,{\sc iii} 4552~\AA\,and these are given in Fig.~3.

Distance estimates have been calculated using the spectroscopic parallax equations
given in Diplas \& Savage \cite{dip94}. Absolute magnitudes have been estimated from the
spectral type and these estimates are based on data from Schmidt-Kaler (1982) and have 
random errors of some 25 per cent. Reddenings have been estimated using the observed ($B-V$) colour 
in combination with the colours and spectral types from stars taken from Wegner (1994). We note that 
especially for mid O-type stars, the absolute magnitudes from Schmidt-Kaler (1982) are systematically fainter 
than in, for example, Wegner (2000), although show better agreement with the estimates from 
Vacca, Garmany \& Shull (1996). The Schmidt-Kaler (1982) magnitudes are chosen as these 
cover all of our spectral types with the exception of WR stars, whose distances we have 
obtained from the literature. We find that agreement
between the Hipparcos stellar parallax distance (obtained from the {\sc simbad} database, operated at CDS,
Strasbourg, France) and our spectroscopic distances are better than 20~percent for luminosity class V objects. For
luminosity class III objects the stellar and spectroscopic distance estimates can differ by as much as a factor of
two. There are no luminosity class I or II objects in our sample within 300~pc and so this relationship cannot be tested for the
highest luminosity class objects. We have decided to adopt the spectroscopic distance estimates for all the stars to
maintain consistency.

\begin{figure}
\epsfig{file=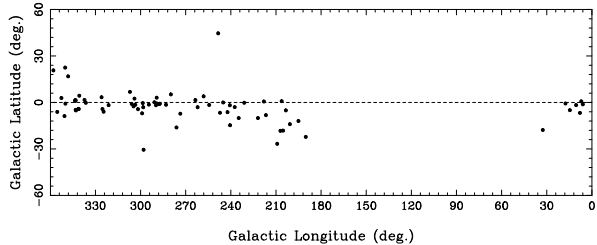, height=90mm, angle=-90}
\caption[]{Galactic location of the stellar sample in latitude and longitude.}
\label{f_gal_latvlong}
\end{figure}

\begin{figure}
\epsfig{file=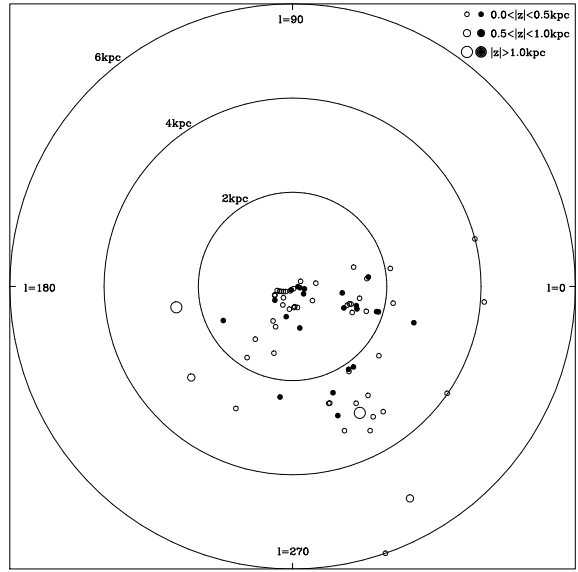, height=90mm, angle=-90}
\caption[]{Heliocentric Galactic location of the stellar sample projected onto
the Galactic plane. Filled circles are stars above the plane of the Galaxy and
open circles are stars below the plane of the Galaxy.} 
\label{f_gal_location}
\end{figure}

In Table~\ref{t_stars} we also list
the height, $z$, of the star relative to the plane of the Galaxy 
($z$=$d_{\rm spec}$ $\times$ sin($b$)).
In Figs.~\ref{f_gal_latvlong} and \ref{f_gal_location} we present a plot
of the Galactic location of the field stars. The majority of the stars lie 
close to the Galactic plane and have Galactic longitudes in the range $l$=270--360$^{\circ}$. 

\begin{table*}
\begin{center}
\small
\caption[]{The stellar sample and basic parameters. Distances for the
  majority of the stars were calculated spectroscopically and errors
  are of the order of 25 per cent, excluding systematic errors (see
  Sect. 2). For some objects the distance was taken from the literature, 
  references to which are given at the bottom of the table.}
\label{t_stars}
\begin{tabular}{lrrrrrrrr}
\hline
    Star    &     $l$~~&    $b$~~ &   Sp. type              & $m_{v}$ & $(B-V)$  & ~$E(B-V)$  &    $d$     &  $z$   \\
            &   (deg.) &   (deg.) &                         &  (mag.) &  (mag.)  &   (mag.)   &    (pc)    &  (pc)  \\
\hline

HD\,164794  &    6.01  &   -1.21  &               O4V f,el  &   5.93  &   0.00  &   0.00  &  1585$^{1}$  &   -33  \\
HD\,163800  &    7.05  &    0.69  &                O7III f  &   7.01  &   0.31  &   0.60  &  1621        &    19  \\
HD\,170235  &    7.95  &   -6.73  &               B2V n,el  &   6.59  &   0.07  &   0.28  &   501        &   -58  \\
HD\,167264  &   10.46  &   -1.74  &             B0.5Ia/Iab  &   5.35  &   0.00  &   0.20  &  2109        &   -64  \\
HD\,171432  &   14.62  &   -4.98  &               B1/B2Iab  &   7.11  &   0.15  &   0.32  &  4014        &  -348  \\
HD\,169454  &   17.54  &   -0.67  &                B1Ia el  &   6.65  &   0.74  &   0.93  &  1359        &   -15  \\
HD\,188294  &   32.65  &  -17.77  &                  B8V d  &   6.44  &  -0.02  &   0.09  &   212        &   -64  \\
HD\, 30677  &  190.18  &  -22.22  &             B1II/III n  &   6.84  &  -0.02  &   0.17  &  2707        & -1023  \\
HD\, 36861  &  195.05  &  -11.99  &             O8III f el  &   3.30  &   0.18  &   0.45  &   347        &   -72  \\
HD\, 37490  &  200.73  &  -14.03  &               B3III el  &   4.57  &  -0.11  &   0.05  &   304        &   -73  \\
HD\, 43285  &  203.42  &   -5.13  &                 B6V el  &   6.05  &  -0.11  &   0.03  &   265        &   -23  \\
HD\, 36646  &  205.31  &  -18.17  &                  B4V n  &   6.63  &  -0.12  &   0.04  &   437        &  -136  \\
HD\, 48099  &  206.21  &    0.80  &               O7V f,el  &   6.37  &  -0.08  &   0.21  &  1640        &    22  \\
HD\, 37055  &  207.04  &  -18.34  &                 B3IV v  &   6.41  &  -0.12  &   0.04  &   449        &  -141  \\
HD\, 33328  &  209.14  &  -26.69  &         B2IV n,el,BCep  &   4.25  &  -0.18  &   0.01  &   251        &  -112  \\
HD\, 45725  &  216.66  &   -8.21  &                 B3V el  &   4.60  &  -0.10  &   0.08  &   184        &   -26  \\
HD\, 52918  &  218.01  &    0.61  &               B1V bCep  &   4.99  &  -0.20  &   0.03  &   477        &     5  \\
HD\, 46185  &  221.97  &  -10.08  &               B2/B3II?  &   6.79  &  -0.16  &  -0.02  &  2937        &  -514  \\
HD\, 58343  &  231.09  &   -0.21  &               B2V n,el  &   5.20  &  -0.04  &   0.17  &   309        &    -1  \\
HD\, 50896  &  234.76  &  -10.08  &                 WN5 wr  &   6.74  &  -0.06  &   0.17  &  1393$^{2}$  &  -243  \\
HD\, 58978  &  237.41  &   -3.00  &                B1II el  &   5.61  &  -0.13  &   0.06  &  1796        &   -93  \\
HD\, 49131  &  240.50  &  -14.73  &                  B2III  &   5.80  &  -0.19  &   0.00  &   870        &  -221  \\
HD\, 61429  &  240.65  &   -1.84  &                 B8IV v  &   4.70  &  -0.11  &  -0.01  &   108        &    -3  \\
HD\, 58377  &  242.20  &   -6.26  &                   B5IV  &   6.81  &  -0.17  &  -0.02  &   448        &   -48  \\
HD\, 64972  &  245.09  &   -0.02  &           B8 p (B6 II)  &   7.18  &  -0.11  &  -0.05  &  2858        &     0  \\
HD\, 60498  &  247.20  &   -6.64  &                  B4III  &   7.35  &  -0.12  &   0.03  &   936        &  -108  \\
HD\, 90882  &  248.41  &   44.59  &                  B9.5V  &   5.18  &  -0.04  &   0.01  &   112        &    78  \\
HD\, 68761  &  254.37  &   -1.62  &                B0.5III  &   6.53  &  -0.09  &   0.13  &  1473        &   -41  \\
HD\, 74966  &  258.08  &    3.93  &                   B4IV  &   7.43  &  -0.14  &   0.01  &   658        &    45  \\
HD\, 72067  &  262.08  &   -3.08  &               B2V n,el  &   5.83  &  -0.16  &   0.05  &   488        &   -26  \\
HD\, 76341  &  263.54  &    1.52  &                O9Ib el  &   7.17  &   0.22  &   0.49  &  2365        &    62  \\
HD\, 76131  &  273.58  &   -7.27  &                  B6III  &   6.69  &  -0.05  &   0.08  &   453        &   -57  \\
HD\, 67536  &  276.11  &  -16.14  &                B2.5V n  &   6.24  &  -0.09  &   0.12  &   450        &  -125  \\
HD\, 89587  &  279.83  &    5.19  &                  B3III  &   6.87  &  -0.13  &   0.03  &   900        &    81  \\
HD\, 88661  &  283.08  &   -1.48  &                B2IV el  &   5.76  &  -0.12  &   0.08  &   458        &   -11  \\
HD\, 92740  &  287.17  &   -0.85  &                WN7A wr  &   6.42  &   0.08  &   0.08  &  2600$^{1}$  &   -38  \\
HD\, 93131  &  287.67  &   -1.08  &                WN6A wr  &   6.48  &  -0.03  &   0.08  &  2600$^{1}$  &   -49  \\
HD\, 94910  &  289.18  &   -0.69  &               B2 pe wr  &   7.09  &   0.49  &   0.08  &  6000$^{3}$  &   -72  \\
HD\, 96917  &  289.28  &    3.06  &               O8.5Ib v  &   7.11  &   0.05  &   0.32  &  2910        &   155  \\
HD\, 94963  &  289.76  &   -1.81  &           O6.5III f,el  &   7.14  &  -0.10  &   0.19  &  3257        &  -102  \\
HD\, 97253  &  290.79  &    0.09  &        O5.5III f,cl,el  &   7.11  &   0.15  &   0.45  &  2415        &     3  \\
HD\,100841  &  294.47  &   -1.40  &                B9III d  &   3.12  &  -0.04  &   0.03  &    52        &    -1  \\
HD\, 29138  &  297.99  &  -30.54  &                  B1Iab  &   7.19  &  -0.09  &   0.10  &  3530        & -1793  \\
HD\,105071  &  298.24  &   -3.09  &                B6Ia/Ib  &   6.31  &   0.13  &   0.19  &  3482        &  -187  \\
HD\,106068  &  298.51  &   -0.41  &                B8Ia/Ib  &   5.95  &   0.23  &   0.26  &  2824        &   -20  \\
HD\,105056  &  298.94  &   -7.05  &          ON9.7Ia pv,el  &   7.42  &   0.09  &   0.26  &  5184$^{1}$  &  -636  \\
HD\,109867  &  301.71  &   -4.35  &                 B1Ia v  &   6.26  &   0.00  &   0.19  &  3264        &  -247  \\
HD\,112272  &  303.49  &   -1.50  &               B0.5Ia v  &   7.39  &   0.64  &   0.84  &  2168        &   -56  \\
HD\,112842  &  304.06  &    2.48  &                   B4IV  &   7.08  &   0.15  &   0.25  &  2126        &    91  \\
HD\,113904  &  304.68  &   -2.49  &            WC5/B0Ia wr  &   5.69  &  -0.11  &   0.11  &  2817        &  -122  \\
HD\,115363  &  305.88  &   -0.97  &                   B1Ia  &   7.82  &   0.50  &   0.69  &  3282        &   -55  \\
HD\,115842  &  307.08  &    6.83  &              B0.5Ia el  &   6.04  &   0.21  &   0.41  &  2157        &   256  \\
HD\,136239  &  321.23  &   -1.75  &               B1.5Ia v  &   7.87  &   0.77  &   0.94  &  2350        &   -71  \\
HD\,143448  &  324.55  &   -5.97  &                B3IV el  &   7.10  &   0.00  &   0.16  &   520        &   -54  \\
HD\,142758  &  325.31  &   -4.28  &               B1.5Ia v  &   7.08  &   0.14  &   0.31  &  4000        &  -298  \\
HD\,137753  &  325.93  &    3.34  &                   B7IV  &   6.70  &  -0.03  &   0.09  &   282        &    16  \\
HD\,148937  &  336.37  &   -0.22  &             O6.5 pn,el  &   6.77  &   0.24  &   0.09  &  1380$^{1}$  &    -5  \\
HD\,148379  &  337.25  &    1.58  &              B1.5Ia el  &   5.36  &   0.44  &   0.61  &  1180        &    32  \\
HD\,148688  &  340.72  &    4.35  &                B1Ia el  &   5.33  &   0.27  &   0.46  &  1452        &   110  \\
HD\,154811  &  341.06  &   -4.22  &                OC9.7Ib  &   6.93  &   0.40  &   0.46  &  1230$^{4}$  &   -90  \\
\hline
\end{tabular}
\normalsize
\end{center}
\end{table*}
 
\begin{table*}
\addtocounter{table}{-1}
\begin{center}
\small
\caption[]{$ctd.$}
\begin{tabular}{lrrrrrrrr}
\hline
    Star    &     $l$~~&    $b$~~ &   Sp. type              & $m_{v}$ & $(B-V)$  & ~$E(B-V)$ &    $d$     &  $z$   \\
            &   (deg.) &   (deg.) &                        &  (mag.) &  (mag.)   &   (mag.)  &    (pc)         &  (pc)  \\
\hline

HD\,154873  &  341.35  &   -4.11  &                 B1Ib d  &   6.70  &   0.28  &   0.47  &  1654        &  -118  \\
HD\,156575  &  342.94  &   -5.16  &                B1.5III  &   7.35  &   0.15  &   0.35  &  1267        &  -113  \\
HD\,156385  &  343.16  &   -4.76  &                 WC7 wr  &   6.92  &   0.05  &   0.35  &  1300$^{5}$  &  -107  \\
HD\,151932  &  343.22  &    1.43  &                WN7A wr  &   6.49  &   0.23  &   0.35  &  1413$^{1}$  &    35  \\
HD\,152235  &  343.31  &    1.10  &                B1Ia el  &   6.34  &   0.42  &   0.61  &  1861        &    35  \\
HD\,152003  &  343.33  &    1.41  &              O9.7Iab v  &   6.99  &   0.29  &   0.53  &  2687        &    66  \\
HD\,152270  &  343.49  &    1.16  &                 WC7 wr  &   6.60  &   0.23  &   0.53  &  1900$^{6}$  &    38  \\
HD\,145482  &  348.12  &   16.84  &                    B2V  &   4.57  &  -0.15  &   0.06  &   268        &    77  \\
HD\,157038  &  349.95  &   -0.79  &              B1/B2Ia d  &   6.41  &   0.64  &   0.81  &  1444        &   -19  \\
HD\,143275  &  350.10  &   22.49  &                 B0.2IV  &   2.29  &  -0.09  &   0.14  &   173        &    66  \\
HD\,163745  &  350.56  &   -8.79  &                   B5II  &   6.50  &  -0.09  &  -0.01  &  2189        &  -334  \\
HD\,155806  &  352.59  &    2.87  &               O7.5V el  &   5.61  &  -0.06  &   0.23  &  1064        &    53  \\
HD\,163758  &  355.36  &   -6.10  &               O6.5Ia f  &   7.32  &   0.02  &   0.34  &  4103        &  -436  \\
HD\,148184  &  357.93  &   20.68  &               B2V n,el  &   4.28  &   0.31  &   0.52  &   122        &    43  \\
\hline
\end{tabular}
\begin{itemize}
\item[] Reference codes: 
(1)~Diplas \& Savage \cite{dip94},
(2)~Fruscione et al. \cite{fru94},
(3)~Hoekzema, Lamers \& van Genderen (1993),
(4)~Winkler \cite{win97},
(5)~Conti \& Vacca (1990),
(6)~Conti et al. (1983).
\end{itemize}
\normalsize
\end{center}
\end{table*}

\section{Data analysis}

The spectra presented in this paper are the merged versions of the POP survey 
available on-line at http://www.eso.org/uvespop. No further data processing was 
undertaken. The following sections detail the methods used to analyse the reduced 
data. 

\subsection{The spectra}                                       \label{s_spectra}

Spectra for the three species were normalised in the region of the interstellar lines  
by fitting a low order polynomial to the continuum using the spectrum analysis program 
{\sc dipso} (Howarth et al. 2003). It should be noted that both the Na\,{\sc i} UV doublet 
and the Ti\,{\sc ii} line lie well within the Free Spectral Range of the CCD and so
problems with order-merging will not affect the spectra. The Ca\,{\sc ii} \,K line
lies closer to the edge of an echelle order, although no obvious problems were visible in
the spectra due to order-merging. The online table of wavelength ranges effected 
by flattening artefacts has also been checked, and none of the species studied fall 
in effected regions.

In Fig.~\ref{f_velcom} we present the spectra for Na\,{\sc i} UV, Ti\,{\sc ii} and 
Ca\,{\sc ii} K for each line-of-sight. Note that we have plotted the two lines of the 
Na\,{\sc i} UV doublet on the same figure, 
to allow for an easy comparison, and hence only 
the feature at 3302.37\,\AA \, is in the LSR. 
Therefore one must add $-$55.3~km~s$^{-1}$ to the 
velocity of the line at 
3302.97\,\AA \, in order to correct it to the kinematical LSR rest frame. 
We also plot the extent of the full width half maximum (FWHM) of the stellar line, 
centred on the LSR velocity of the 
star. The LSR velocity and FWHM of the stellar lines were estimated by Gaussian
fitting using the Emission Line Fitting suite 
({\sc elf}) within {\sc dipso}. We generally fitted the stellar lines
of the Si\,{\sc ii} doublet at 3853\,\AA\ or the Si\,{\sc iii} triplet at 4552\,\AA\ 
to determine these values.
In some cases we were unable to estimate the FWHM due to the
stellar lines being merged into the continuum. We therefore
approximated
the FWHM of the stellar line as the $v$sin$i$ value for the star, 
taken from the literature for the following stars;
HD\,148184~(Krelowski, Schmidt \& Snow, 1997), HD\,52918~(Goraya \& Tur, 1988),
HD\,58978~(Mennickent et al., 1994), HD\,67536~(Grady, Bjorkman \& Snow, 1987),
HD\,88661~(Ballereau, Chauville \& Zorec, 1995), HD\,45725~(Slettebak, 1982), 
HD\,143275~(Hurwitz \& Bowyer, 1996), HD\,148937 and HD\,164794~(Penny, 1996). 
For the Wolf-Rayet stars we have generally
been unable to detect stellar lines with certainty and have not found $v$sin$i$
values from the literature. Hence we have not plotted the extent of the stellar
line. We have estimated the LSR velocity of these stars from the
hydrogen emission lines and this is listed with the plot for these stars in
Fig.~\ref{f_velcom}.

It can be seen that 
in almost all cases the stellar contamination of the interstellar line is 
minimal, since the stellar feature is sufficiently
broad to merge into the
continuum or has been easily removed through the normalisation process. 
In cases where some stellar contribution is evident, this was fitted
using {\sc elf} and the column density was calculated 
in the {\sc is} fitting process within {\sc dipso}. Hence when we fit the interstellar
components the stellar contribution was removed. It should be
noted that stellar contamination is usually only seen in the Ca\,{\sc ii} K line.
In cases where it was not possible to easily fit the stellar line, i.e. when
it
is of a similar width to the interstellar profile and hence causes significant
contamination, we have not calculated column
densities for the components, but we have listed the velocities of each component
in Table~3.

The high S/N of our data means that we detect interstellar Ca\,{\sc ii} in every one of our spectra 
and Ti\,{\sc ii} in all but one. Even 
in this case, the Ti\,{\sc ii} spectrum towards HD\,90882 does indeed show some
evidence for absorption, although we have chosen not to fit this as it is close to
the noise level. The S/N ratio of this Ti spectrum (190) is somewhat lower than the 
mean for Ti (300), and the star is also one of the closest in our sample 
at $d$=112 pc, with an upper limit to the column denisty of 
$\sim$ 1.2$\times$ 10$^{11}$ cm$^{-2}$ (Sect.\ref{s_ew} and
Sect.\ref{s_colden}). 
We have fewer detections in the Na\,{\sc i} UV lines, with 15 sightlines showing no 
interstellar Na\,{\sc i} absorption and upper limits to the column
density have been calculated along these sightlines.

\begin{figure*}
\epsfig{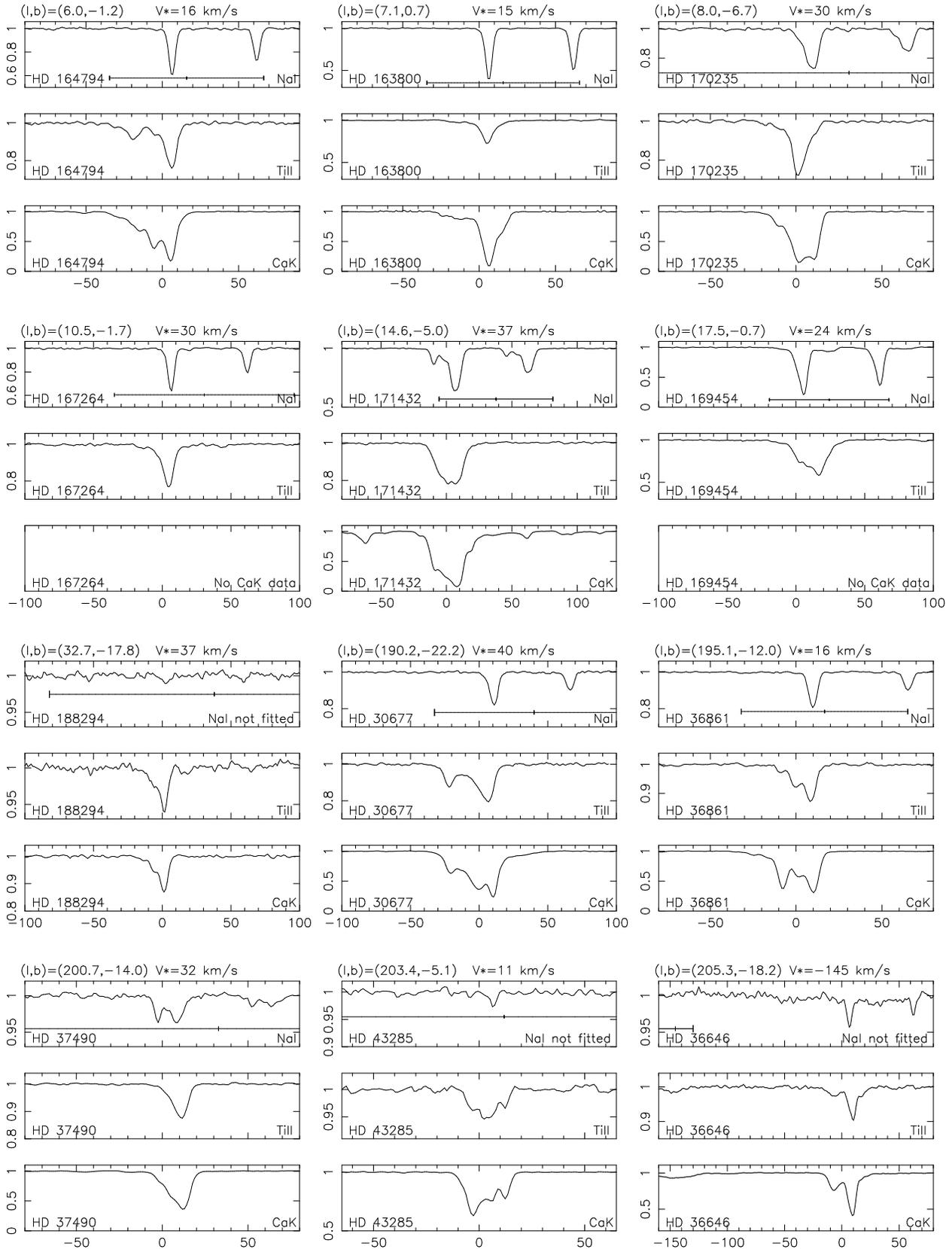}
\caption[]{The interstellar Na\,{\sc i} UV (3302.37\AA), Ti\,{\sc ii} (3383.76\AA) 
and Ca\,{\sc ii} K (3933.66\AA) absorption profiles along each line-of-sight. 
The abscissa is the kinematic 
LSR velocity in km~s$^{-1}$ and the ordinate is the normalised flux. Note that the 
two lines of
Na\,{\sc i} UV have been plotted on the same figure, and a correction of
--55.3~km~s$^{-1}$ should be added to the weaker line at 3302.98\AA \, to 
move it to a LSR
velocity. The horizontal bars indicate the extent of the FWHM of a stellar
line and the stellar velocity is marked in the centre of each line. V$_{*}$ is 
the stellar velocity in the LSR.} 
\label{f_velcom}
\end{figure*}

\begin{figure*}
\setcounter{figure}{2}
\epsfig{file=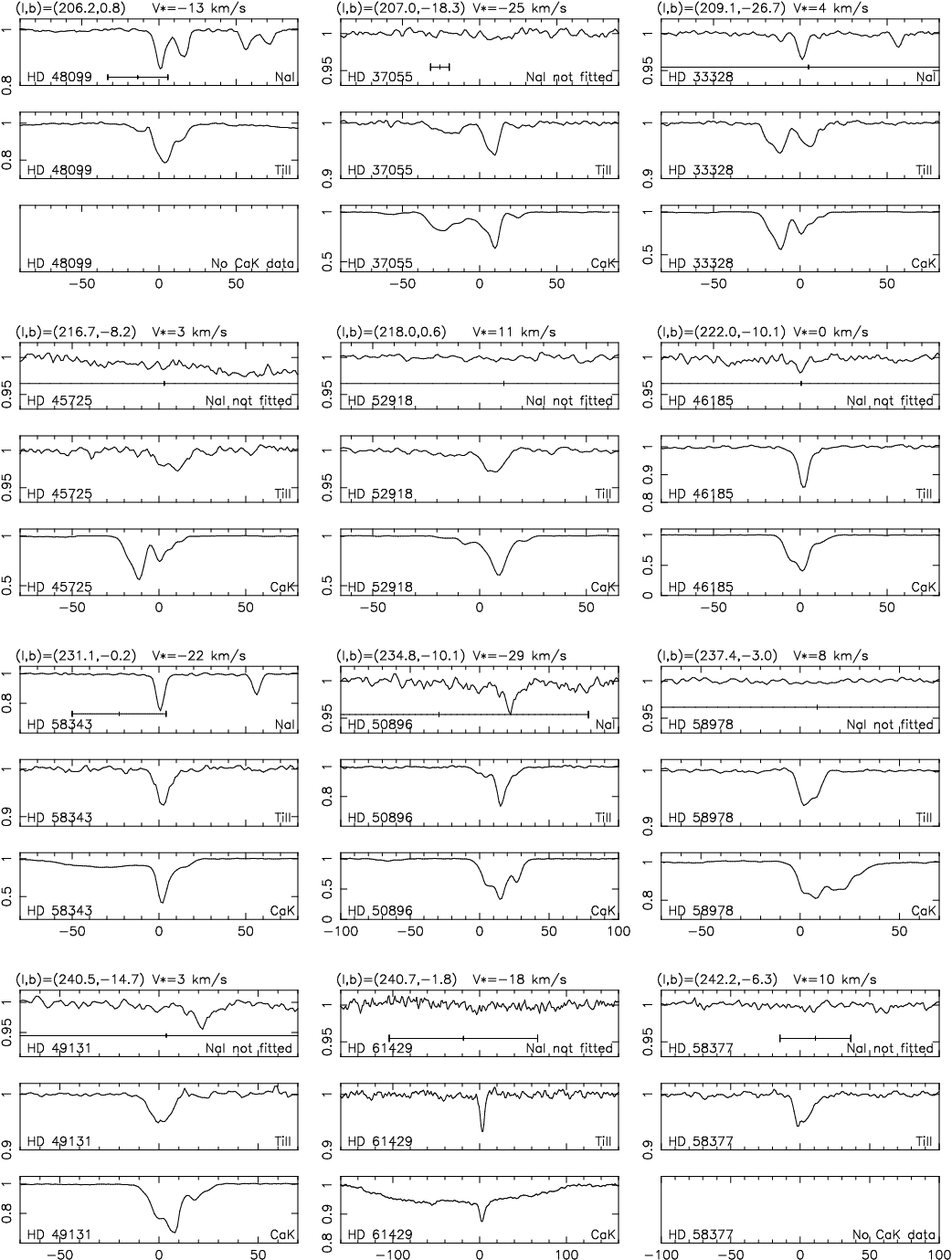, height=220mm, angle=-0}
\caption[]{$ctd.$}
\end{figure*}

\begin{figure*}
\setcounter{figure}{2}
\epsfig{file=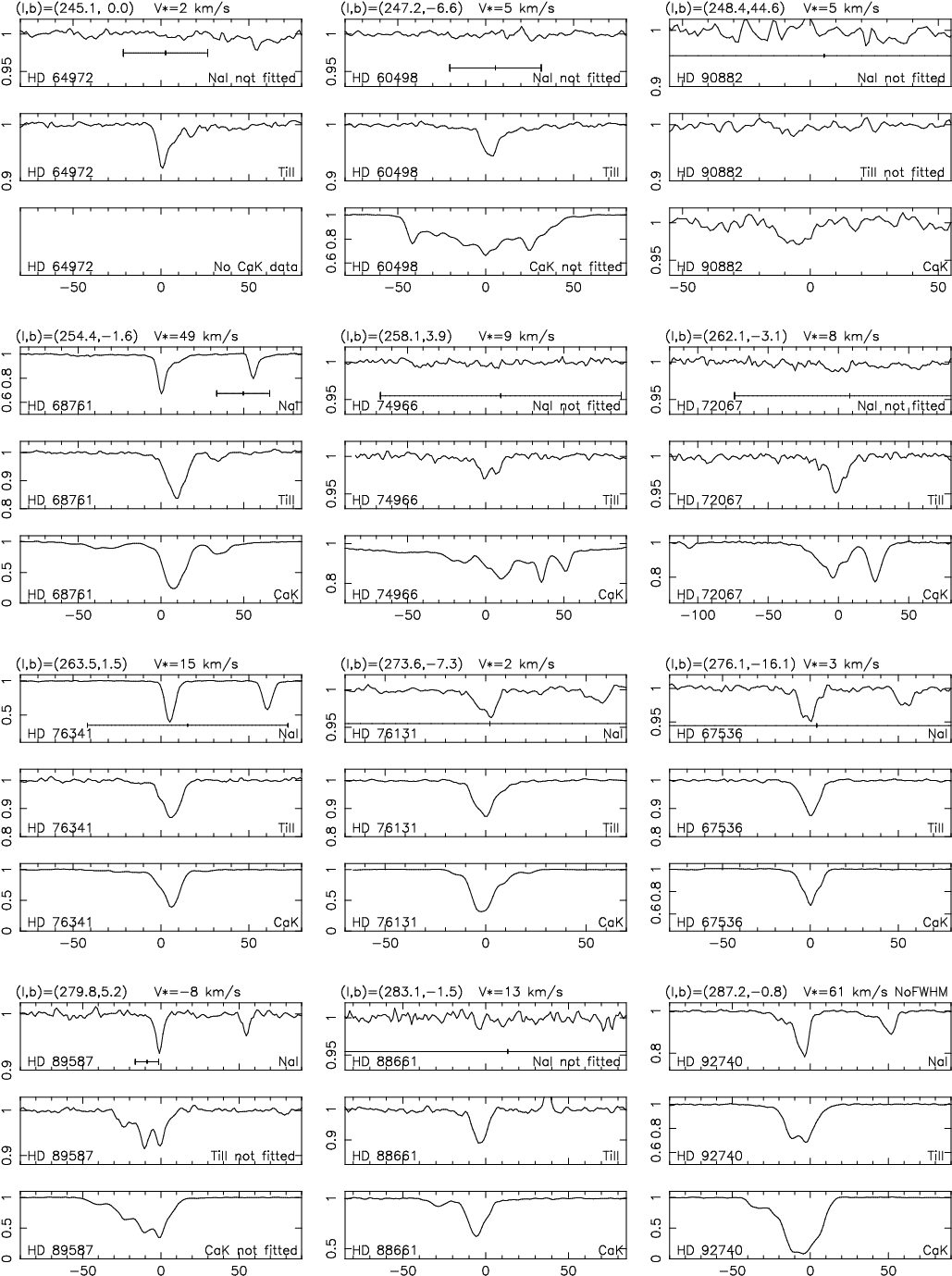, height=220mm, angle=-0}
\caption[]{$ctd.$}
\end{figure*}

\begin{figure*}
\setcounter{figure}{2}
\epsfig{file=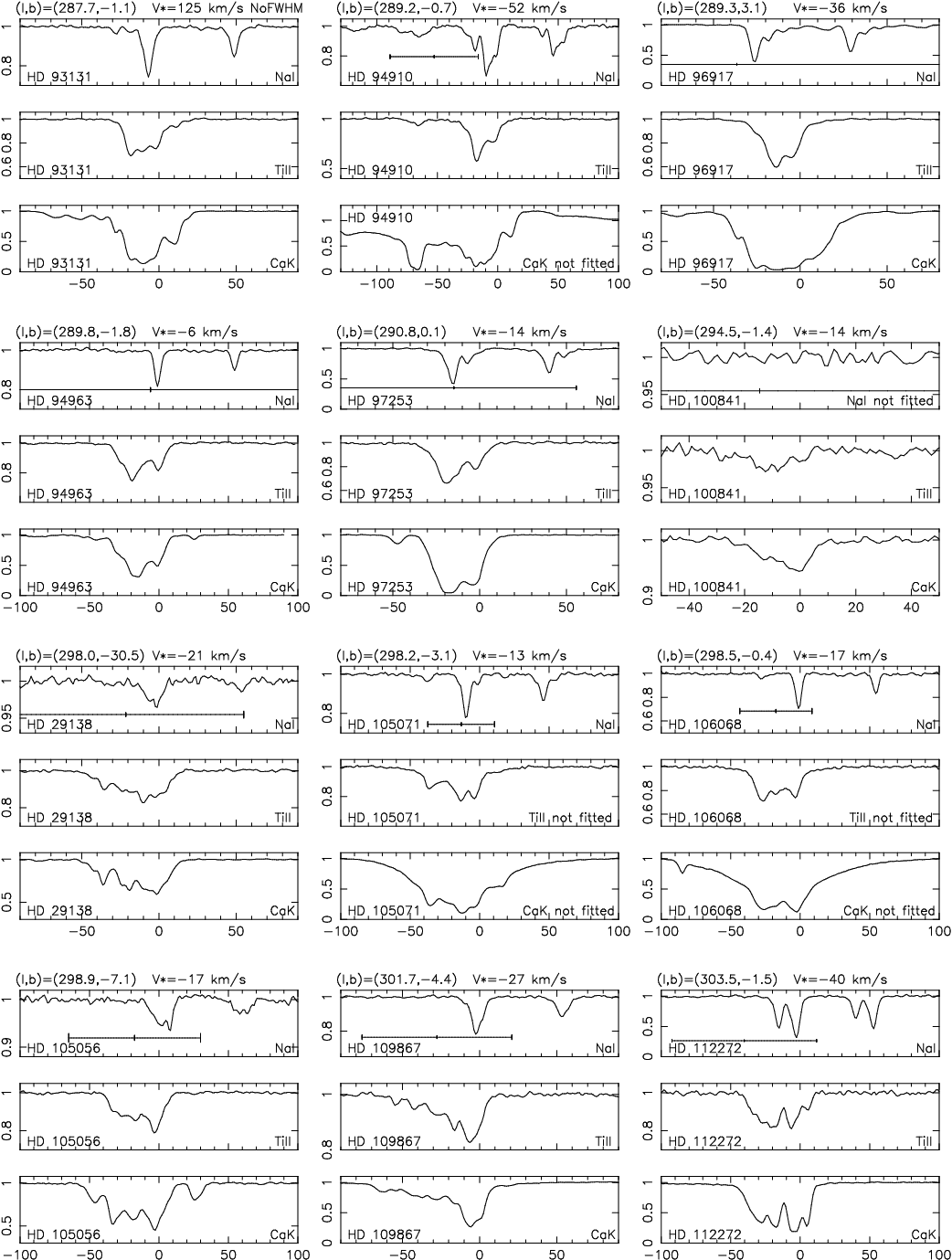, height=220mm, angle=-0}
\caption[]{$ctd.$}
\end{figure*}

\begin{figure*}
\setcounter{figure}{2}
\epsfig{file=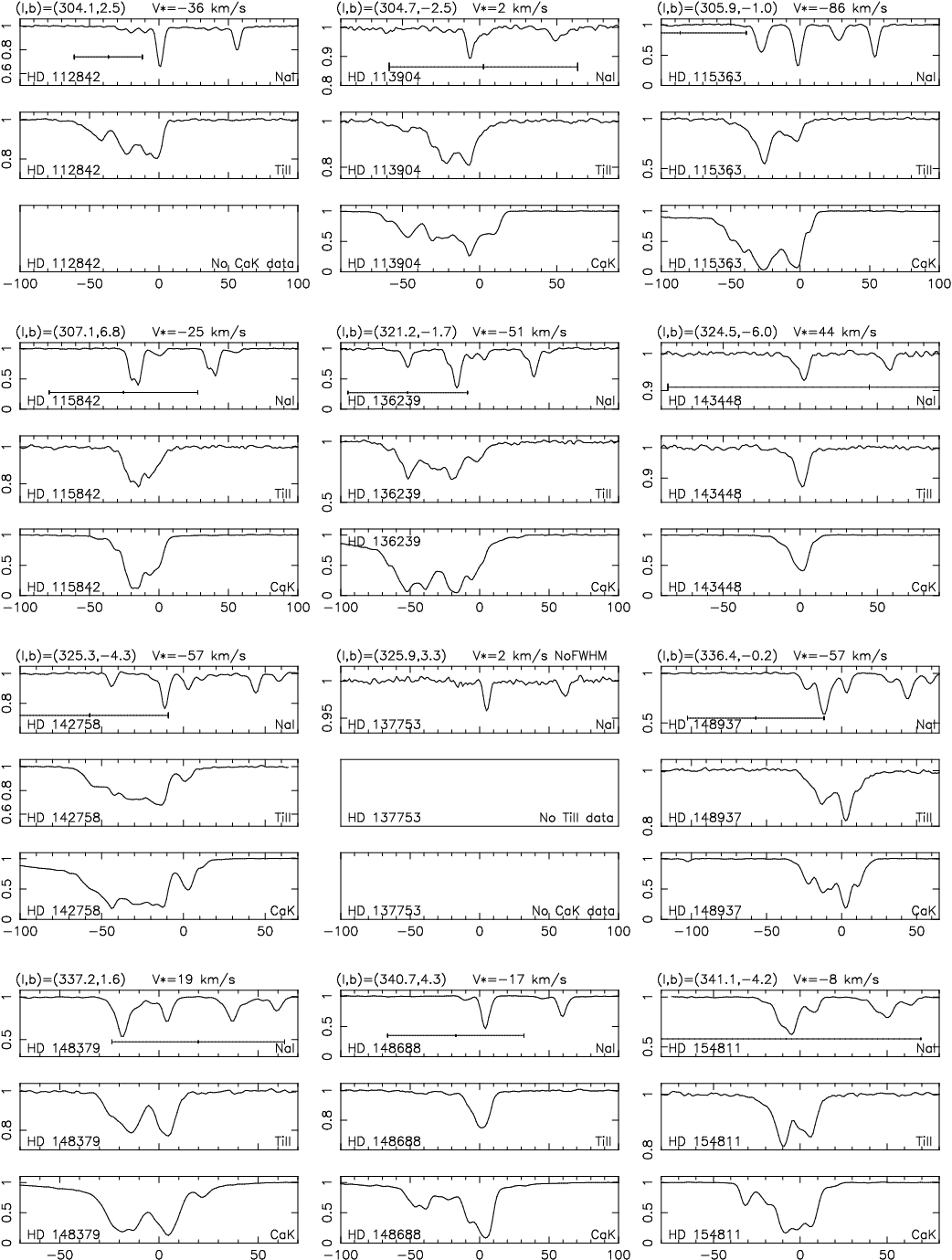, height=220mm, angle=-0}
\caption[]{$ctd.$}
\end{figure*}

\begin{figure*}
\setcounter{figure}{2}
\epsfig{file=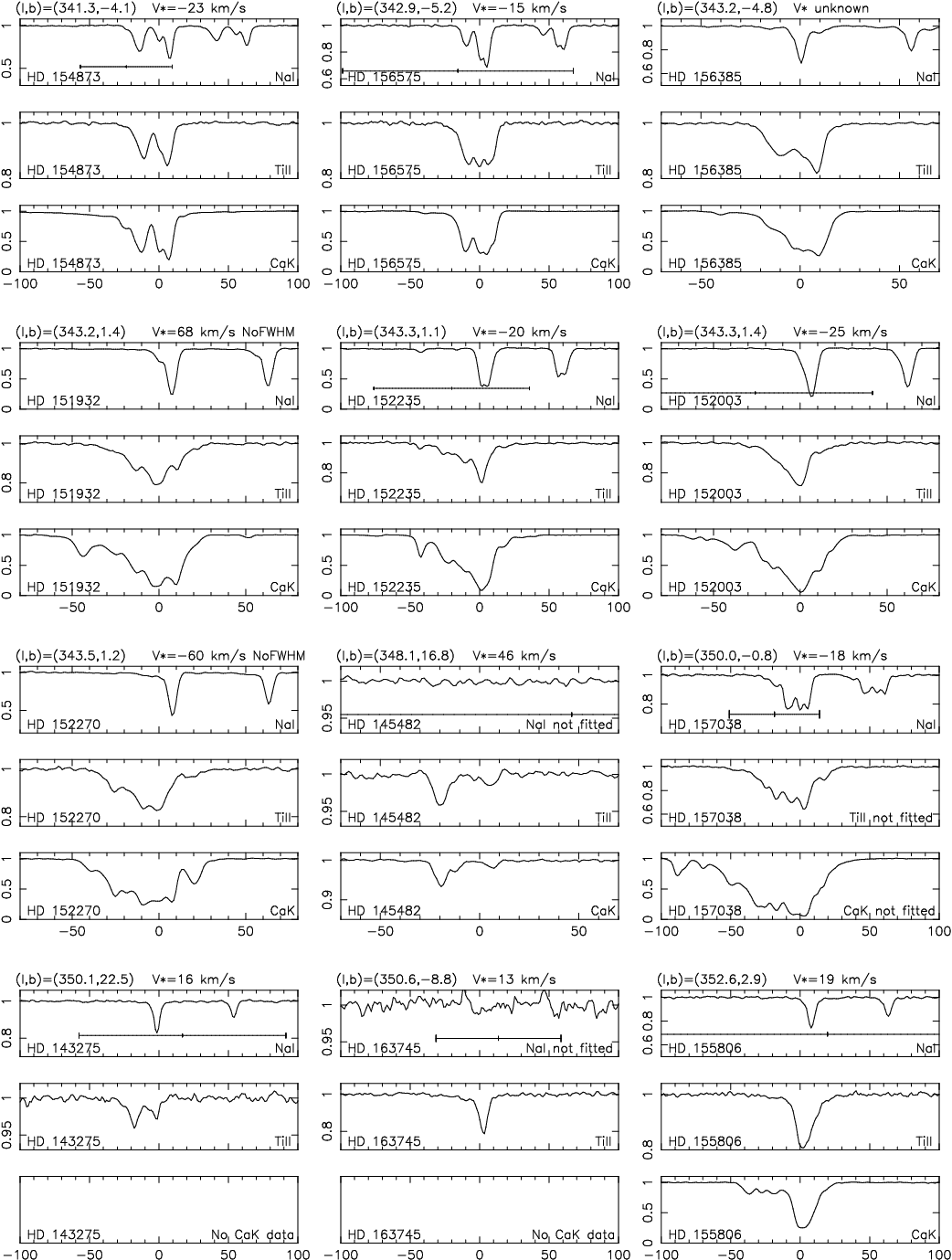, height=220mm, angle=-0}
\caption[]{$ctd.$}
\end{figure*}

\begin{figure*}
\setcounter{figure}{2}
\epsfig{file=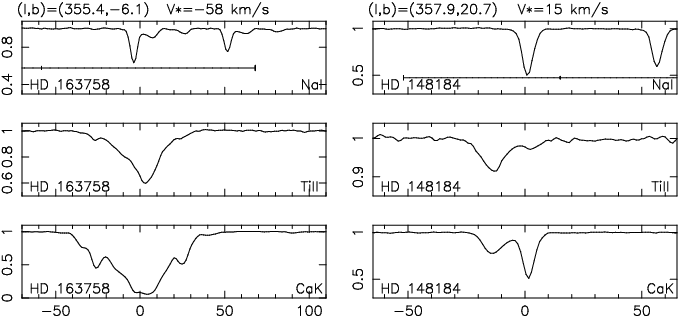, height=55mm, angle=-0}
\caption[]{$ctd.$}
\end{figure*}

\subsection{Equivalent Width Measurements and upper limits}                          \label{s_ew}

The equivalent width ($EW$), FWHM and the 
central velocity ($v$) of each component of each interstellar line were 
measured by fitting Gaussian profiles to the normalised spectra using the 
line fitting suite of programs {\sc elf}.  
The equivalent widths and the velocities for each interstellar component 
are listed in Table~3 along with their errors. The errors in equivalent width
are based on the 3$\sigma$ errors in column density as described in Sect.\ref{s_error_in_com}.
Errors in the velocities were taken from {\sc elf}. In order 
to check their reliability, test data at S/N ratios of 200 and 400 were
created with different $b$-values and the estimated errors compared with 
the known values for the centres of the lines. Agreement was found to be excellent 
between the calculated error and known position of the lines.
 
The determination of upper limits in the case of non-detections is 
non-trivial. Non-detections were estimated using a velocity width and a 
3-$\sigma$ cutoff derived from the S/N ratio. Clearly, the non-detections 
have an unknown velocity width, and often lie in between other detected components.  
Hence the following scheme was adopted. If the undetected component c$_{\rm undet}^{2}$ 
lies in the middle of three detected components in another species 
c$_{\rm detet}^{1}$, c$_{\rm detet}^{2}$, c$_{\rm detet}^{3}$ of velocities 
$v_{\rm detet}^{1}$, $v_{\rm detet}^{2}$, $v_{\rm detet}^{3}$, the lower-velocity 
limit was set at $v_{\rm detet}^{2}$--(($v_{\rm detet}^{2}$--$v_{\rm detet}^{1}$)/2.0) 
and the upper velocity limit at 
$v_{\rm detet}^{2}$+(($v_{\rm detet}^{3}$--$v_{\rm detet}^{2}$)/2.0). In the cases 
where there were more than one species detected in for example component 3, 
the average of these velocities was taken for  $v_{\rm detet}^{3}$. We stress that 
due to the unknown $b$-value of the undetected components, the upper limits may not 
be reliable.
 
\subsection{Column Density Measurements - Component fitting}             \label{s_colden}

The Gaussian velocity dispersion parameter, 
$b$, and the column density, $N$, were initially
estimated based on the FWHM and the EW of each of the individual components as
measured using {\sc elf}. Values of $b$ were determined 
to be 0.6006 $\times$ FWHM
of the profile, corrected for instrumental broadening. The column density was
initially derived from the equation: 

\begin{equation}
N({\rm cm^{-2}})=1.13 \times 10^{17} \times \frac{EW ({\rm m\AA})}{f \times \lambda^{2} 
({\rm \AA)}},
\label{coldensVerg}
\end{equation}

\noindent
where $\lambda$ is the rest wavelength of the line and $f$ is the 
oscillator strength (e.g. Vergely et al. 2000).
Oscillator strengths and rest wavelengths were taken from Morton (2003, 2004).

Eq. \ref{coldensVerg} assumes that the medium is optically thin, which, while a
reasonable approximation for the Na\,{\sc i} UV 
interstellar line doublet at 3302\,\AA, 
and to a lesser extent for Ti\,{\sc ii} 3383\,\AA\ is a poor
assumption for Ca\,{\sc ii} 3933\,\AA. Nevertheless, this method was
useful in gaining initial estimates of the column density for each component. The
{\sc is} suite of programs within {\sc dipso} was then used to refine the column
density, and in some cases the $b$-values of the individual components. In every case
the theoretical interstellar spectrum generated by the {\sc is} program  was in
excellent agreement with the observed one. 

The upper limits to the column denisty for each component was estimated using the upper
limits to the $EW$ (see Sect.~\ref{s_ew}) and Eq.~\ref{coldensVerg} and are given in Table~3.

\subsection{Column Density Measurements - Apparent optical depth method}                         \label{s_AOD}

We have also calculated the total column density by the apparent optical depth
(AOD) method. 
In this, the observed interstellar line profiles are converted
first to AOD and then to apparent column density by integrating across the line
profile. The method and its accuracy are described in detail in Savage \&
Sembach \cite{sav91}. In Table~4 we present the total column
density, derived by both component fitting and by the AOD method for each interstellar 
line observed along
each line-of-sight. Where a stellar line was fitted, as
discussed in Section~\ref{s_spectra}, the column density of this line was
subtracted from the total derived by the AOD method.
In Fig.~\ref{f_compareAOD} we present plots comparing the column density 
derived by profile fitting and the AOD method. It can be seen that the scatter in the plots
is minimal especially in the case of Ti and Ca.
We also see that at higher column densities, the profile fitting and AOD methods
are in better agreement for Ti and Ca. It should be noted that in the profile
fitting and the AOD methods we have used the same base-line fit. We note that 
each of the lines in the Na\,{\sc i} doublet were measured independently.

\subsection{Uncertainty in individual cloud column densities}        \label{s_error_in_com}

From the instrumental FWHM resolution, $\Delta$$\lambda$$_{\rm inst}$, and the
S/N ratio of the spectra, we can calculate the smallest equivalent width,
$EW_{\rm lim}$, of an unresolved feature that we would expect to see at the 3$\sigma$ level by the
relationship:

$EW_{\rm
lim}$~=~3$\times$$\Delta$$\lambda$$_{\rm inst}$~(S/N)$^{-1}$.

Using this, 
we find that $EW_{\rm lim}$ for Na\,{\sc i} UV, Ti\,{\sc ii}
and Ca\,{\sc ii} are 0.45~m\AA, 0.39~m\AA~ and 0.33~m\AA, respectively and this is
in reasonable agreement with the smallest interstellar features we have actually measured, 
see Table~3.
Converting the calculated $EW_{\rm lim}$ to a
column density using Eq. \ref{coldensVerg}, we find that we
should be able to observe changes of 0.10~dex, 0.10~dex and 0.05~dex in the
column density of an individual component for Na, Ti and Ca
respectively, where these values have been calculated for a component with an EW of the mode
of the EW's given in Table~3. This agrees with the observations and 
errors estimates for the column densities of individual components given in Table~3.
The error estimates 
in the $b$-values and column densities for each component were derived 
as follows. For each detected component of each species, the column density was 
increased, in steps of one per cent between 1 and 15 per cent and in steps of five between 15 and 100 per cent,
leaving the other components in the best-fit model constant, until the peak residual (in the data$-$model 
fit) exceeded 3$\sigma$. The difference between the column density 
of the fit at the 3$\sigma$ level for this component and the best-fit column density,
was taken to be the uncertainity. An identical procedure was performed
to estimate the $b$-value errors. Where no 3$\sigma$ limit was found,
the error was set to be 95 per cent of the observed profile. This
occurs in the case of weak features where we may be fitting noise. 
Finally, errors in the equivalent widths were derived 
from the errors in the column density measurements using Eq.~\ref{coldensVerg}.

\subsection{Uncertainty in total column densities}        \label{s_error_totals}

In this section we estimate the
uncertainties in the column densities, by comparing the total value
derived from the stronger and weaker lines in the Na and Ca doublets.
In the case of Ca this allows us to take into account all sources of
error, for example in the continuum fit. For the Na lines the baseline 
fit is not independent. In Fig.~\ref{f_totNaIvtotNaI} we plot the total column
density derived from each line of the Na\,{\sc i} UV doublet and, in general, there is
excellent agreement between the total
column density of each line, from which we can deduce the accuracy of our 
base-line fitting
and column density measurements. The few discrepant cases can normally be
accounted for by noise features affecting weak detections, saturation affecting strong
detections, and in one

\begin{figure*}
\setcounter{figure}{3}
\epsfig{file=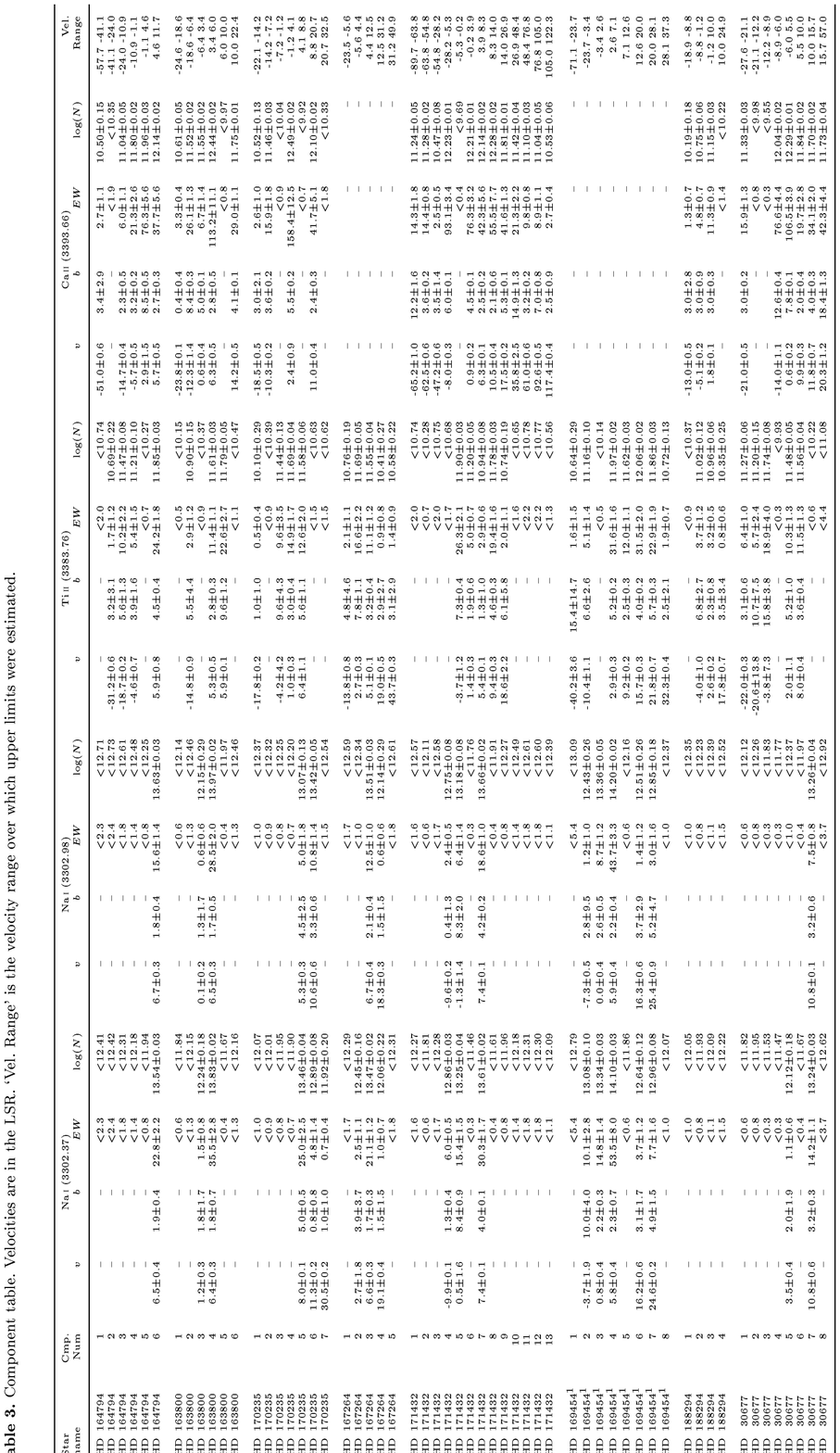, height=245mm, angle=180}
\end{figure*}

\begin{figure*}
\setcounter{figure}{3}
\epsfig{file=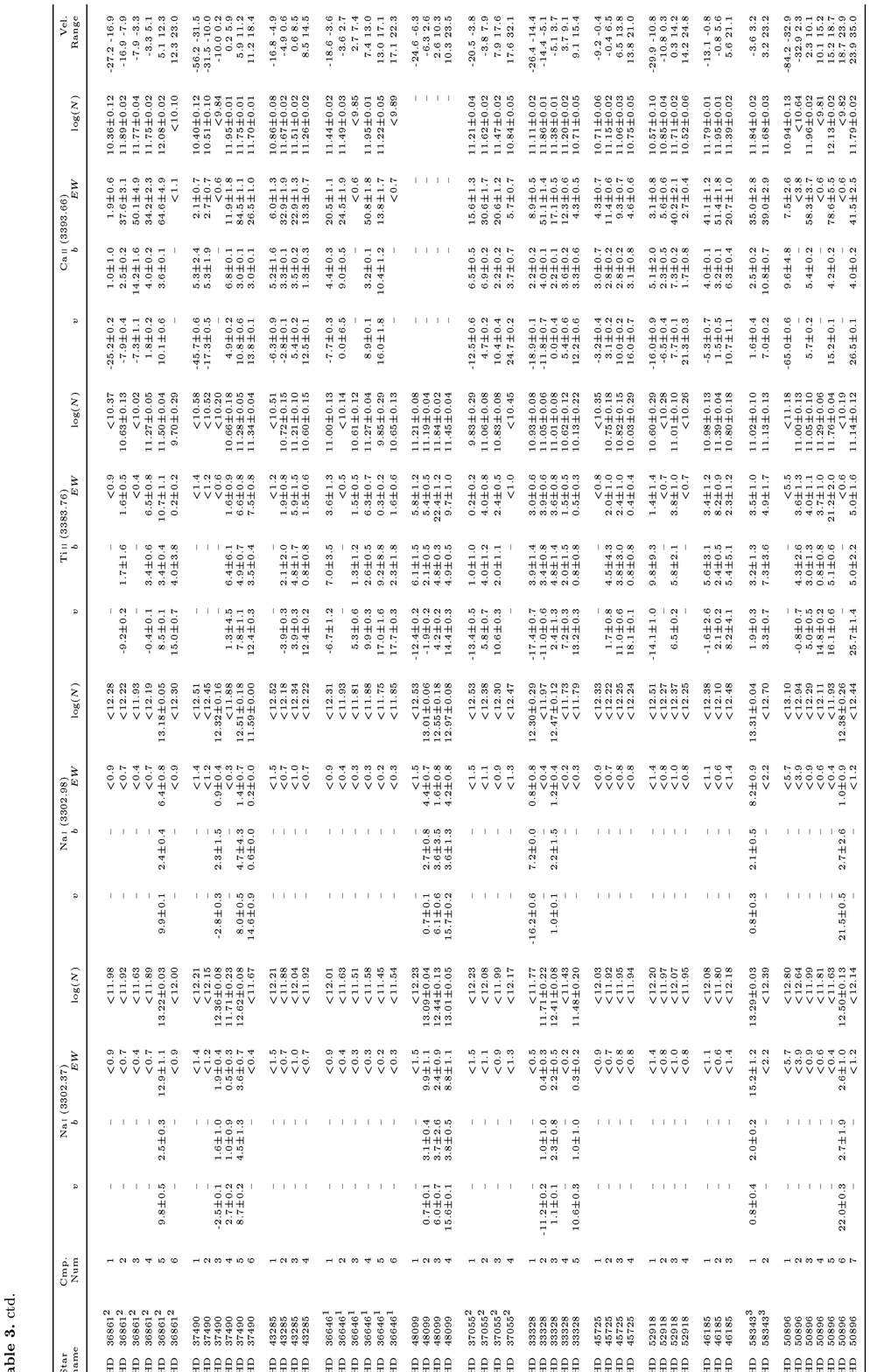,height=245mm, angle=180}
\end{figure*}

\begin{figure*}
\setcounter{figure}{3}
\epsfig{file=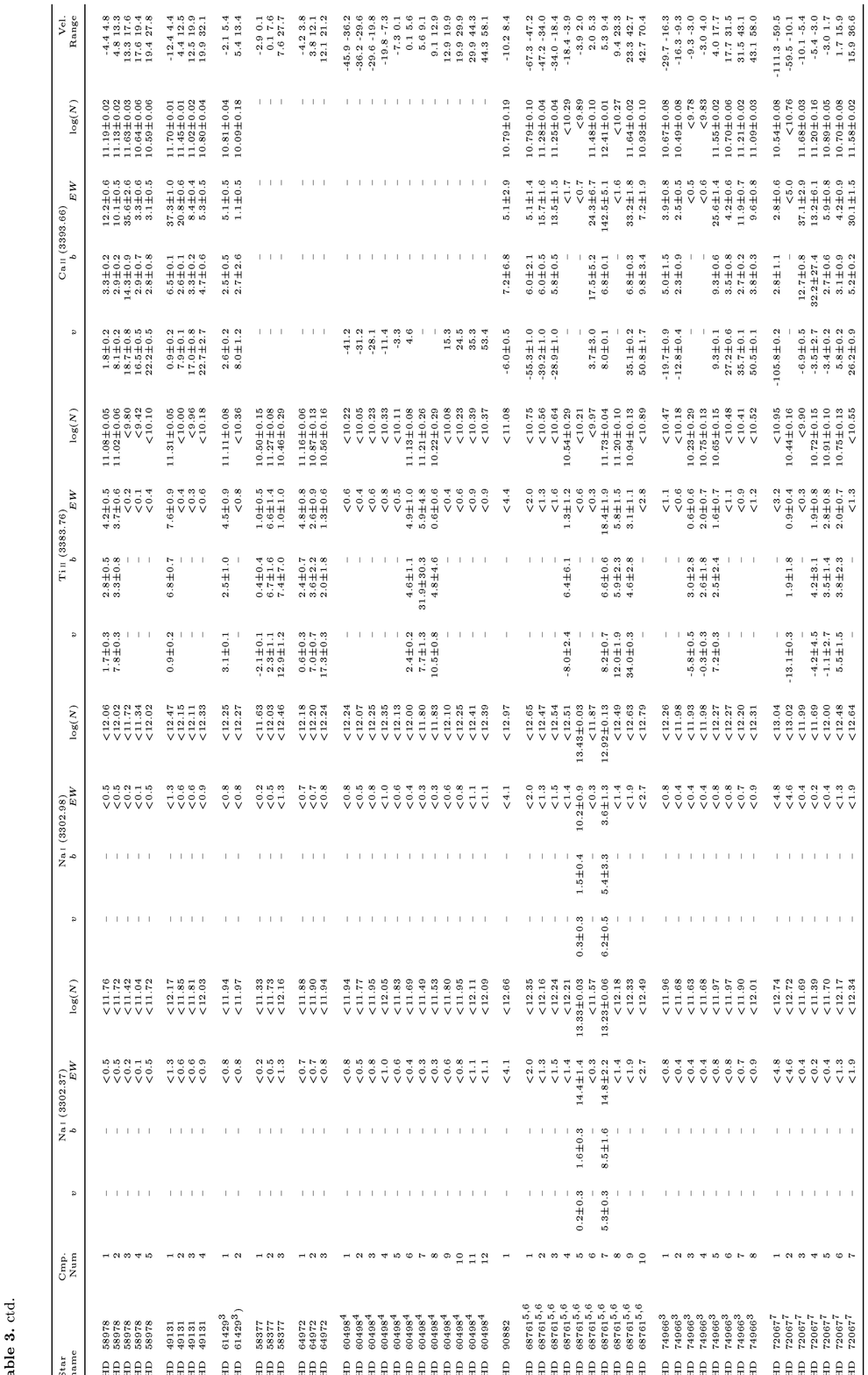,height=245mm, angle=180}
\end{figure*}

\clearpage
\newpage
\begin{figure*}
\setcounter{figure}{3}
\epsfig{file=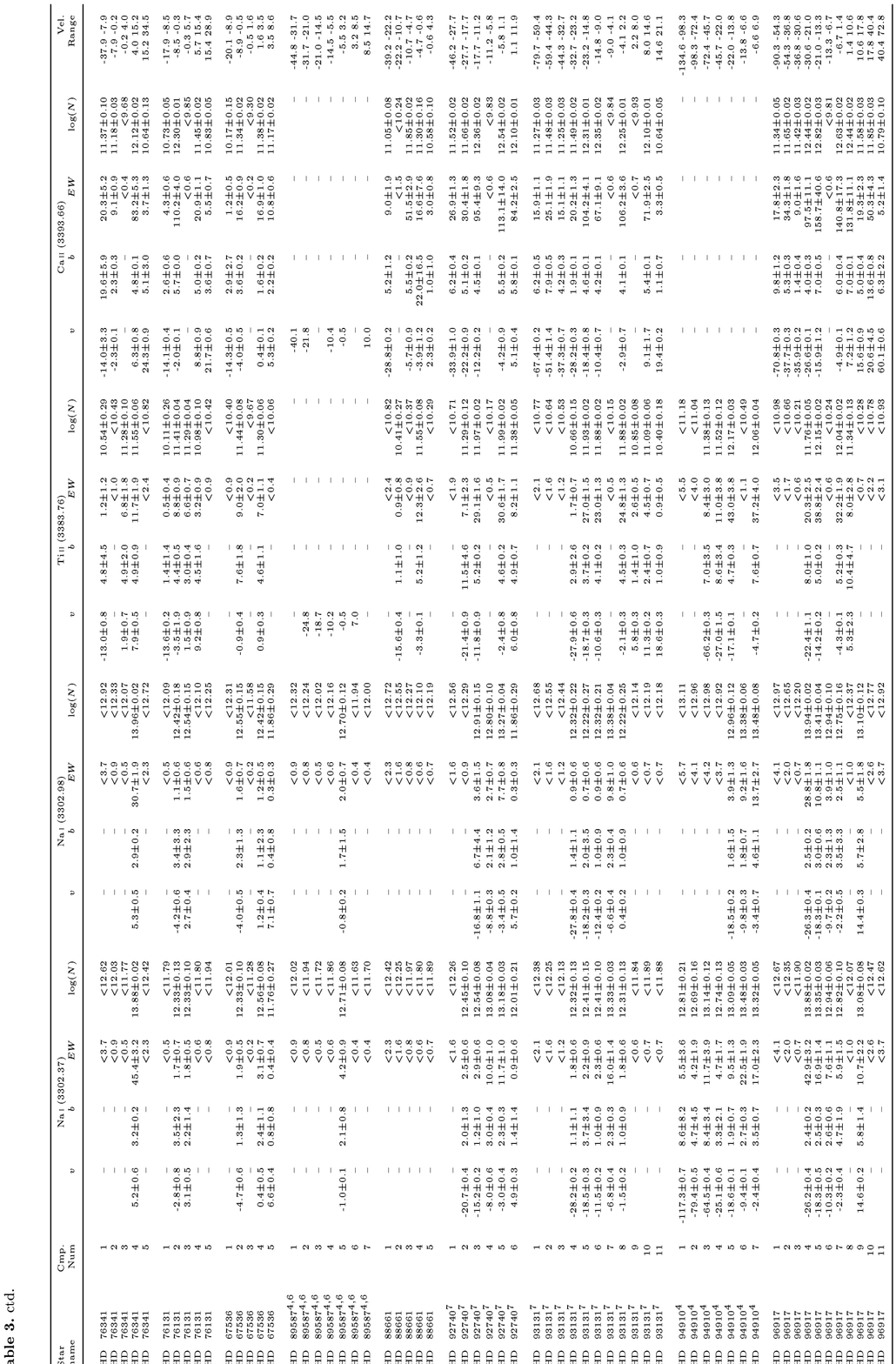,height=245mm, angle=180}
\end{figure*}

\clearpage
\newpage
\begin{figure*}
\setcounter{figure}{3}
\epsfig{file=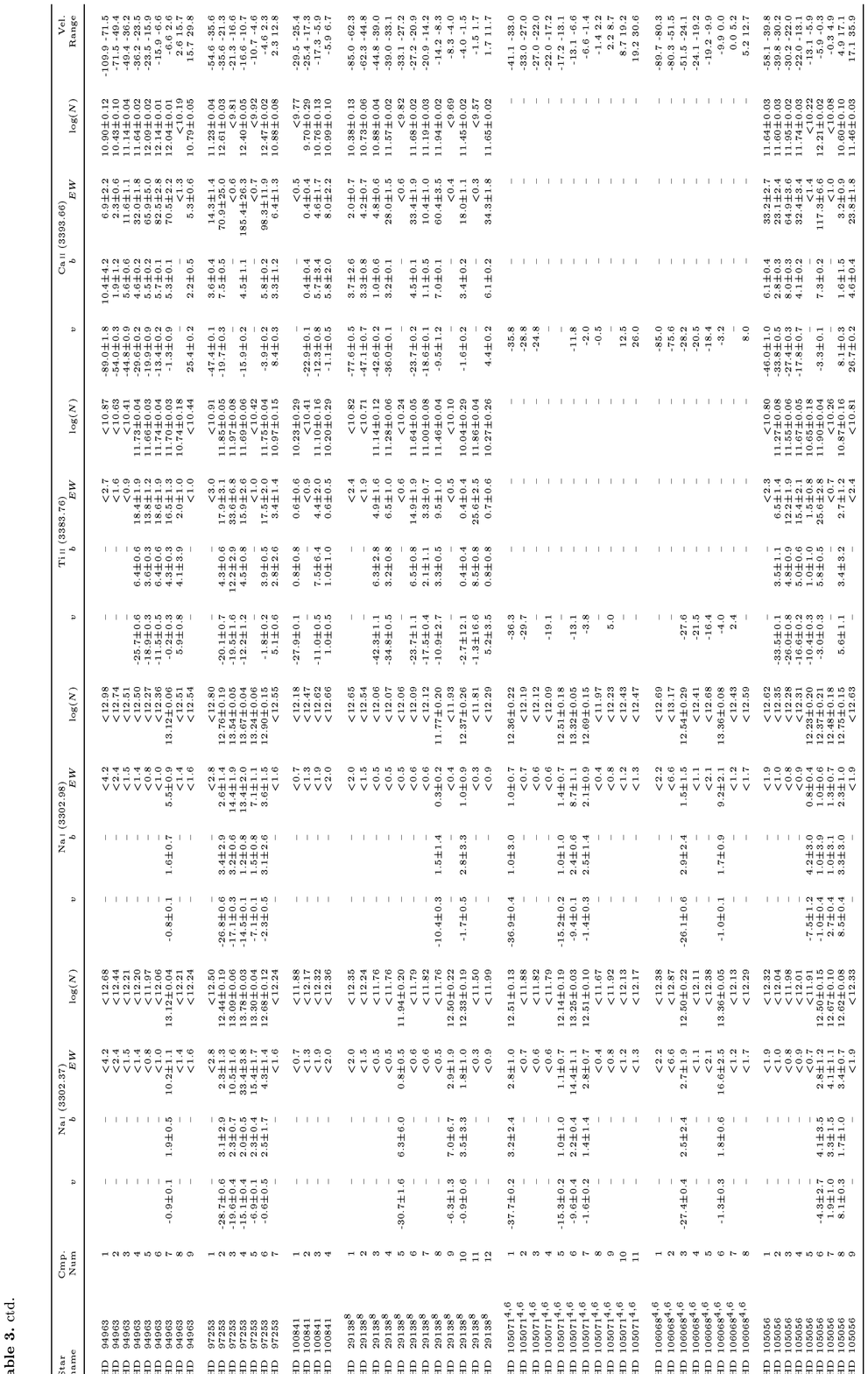,height=245mm, angle=180}
\end{figure*}

\clearpage
\newpage
\begin{figure*}
\setcounter{figure}{3}
\epsfig{file=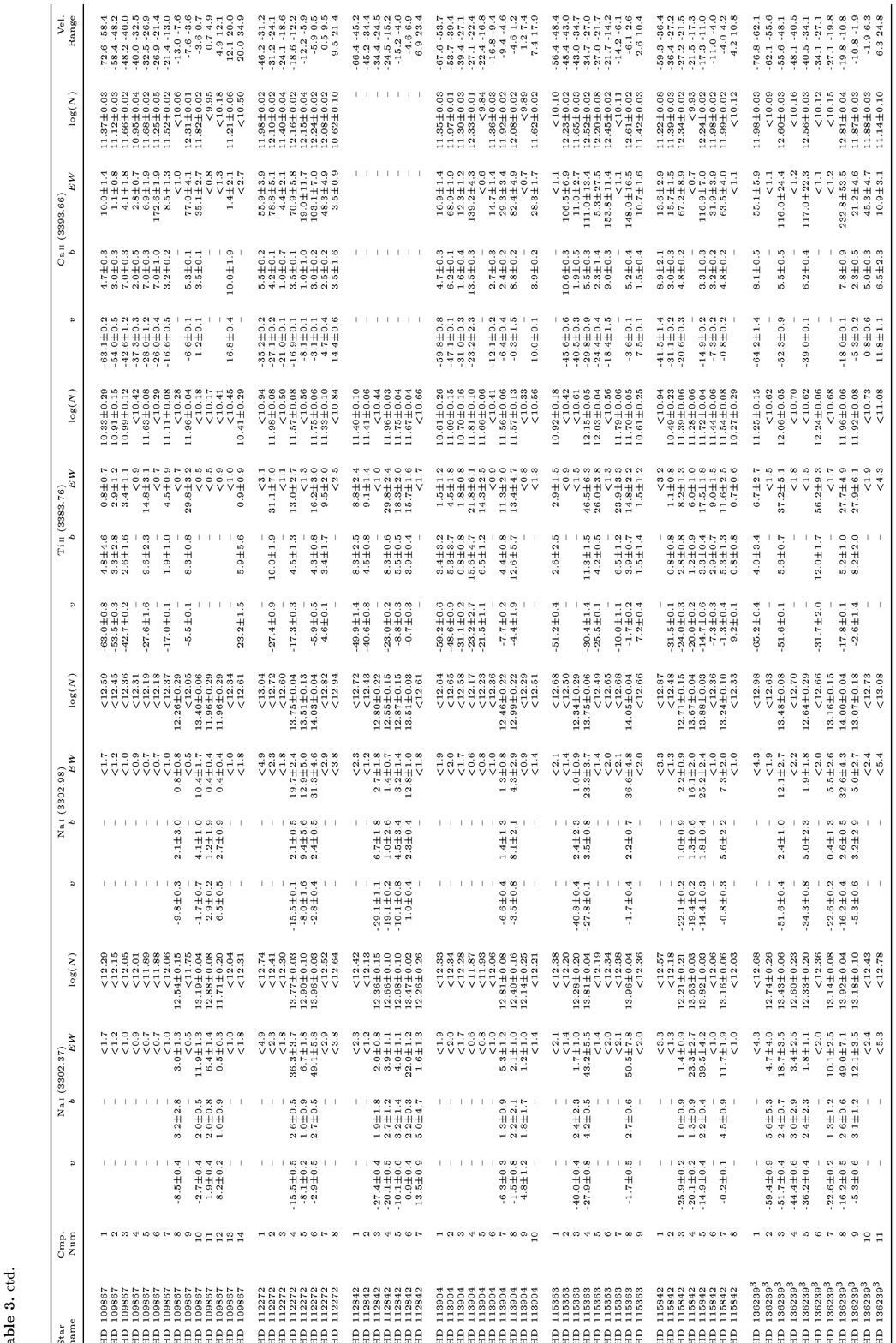,height=245mm, angle=180}
\end{figure*}

\clearpage
\newpage
\begin{figure*}
\setcounter{figure}{3}
\epsfig{file=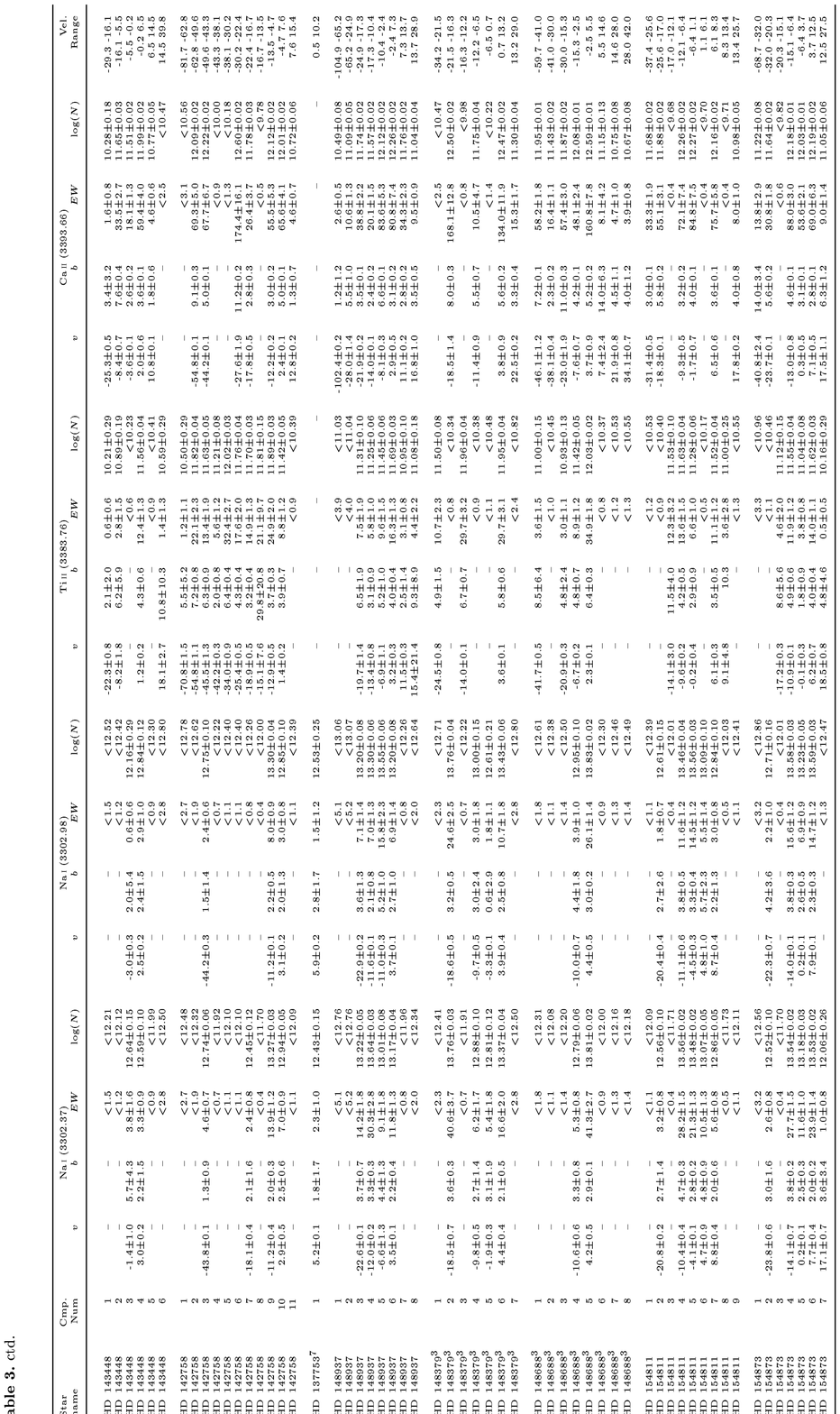,height=245mm, angle=180}
\end{figure*}

\clearpage
\newpage
\begin{figure*}
\setcounter{figure}{3}
\epsfig{file=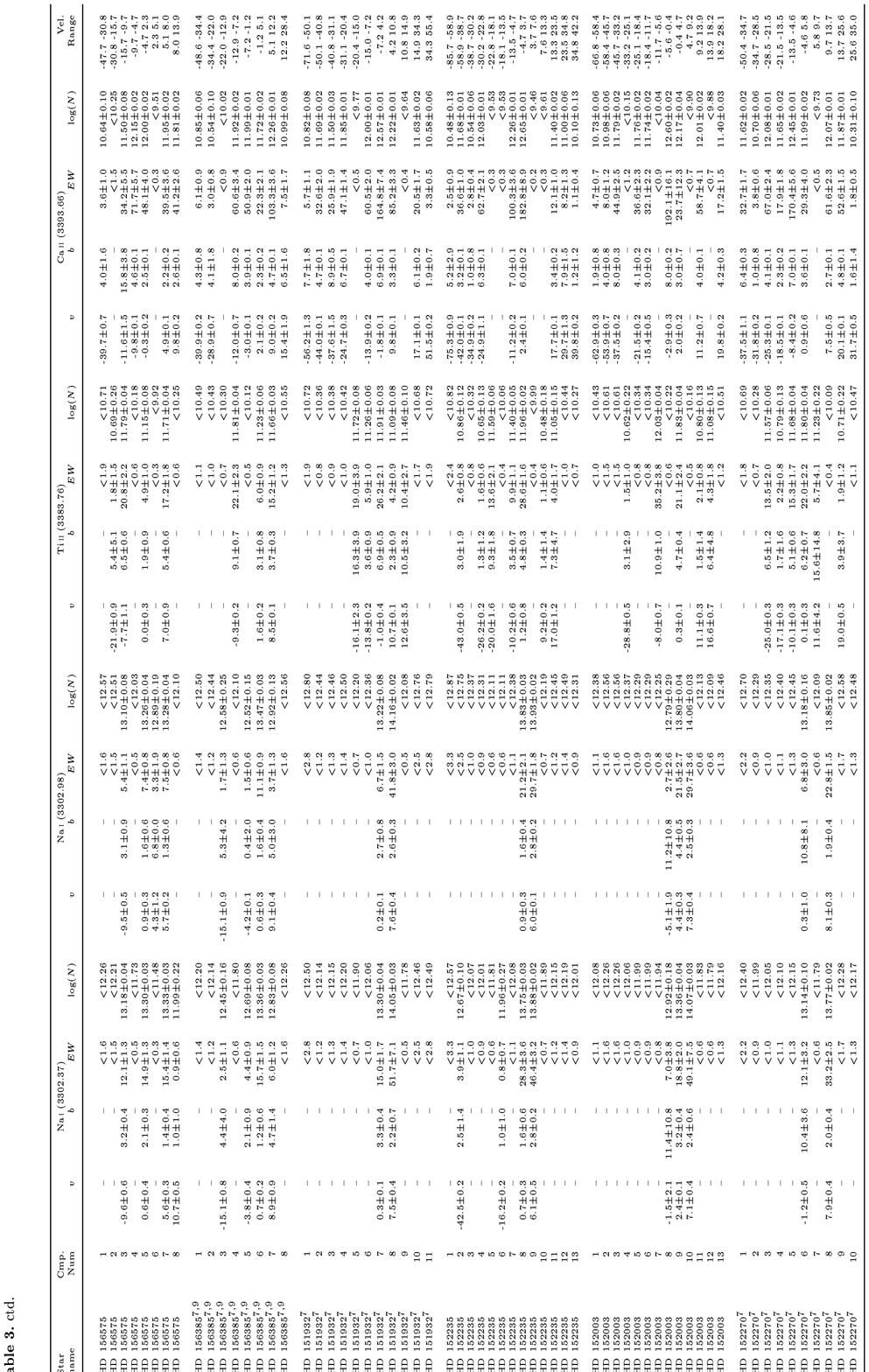,height=245mm, angle=180}
\end{figure*}

\clearpage
\newpage
\begin{figure*}
\setcounter{figure}{3}
\epsfig{file=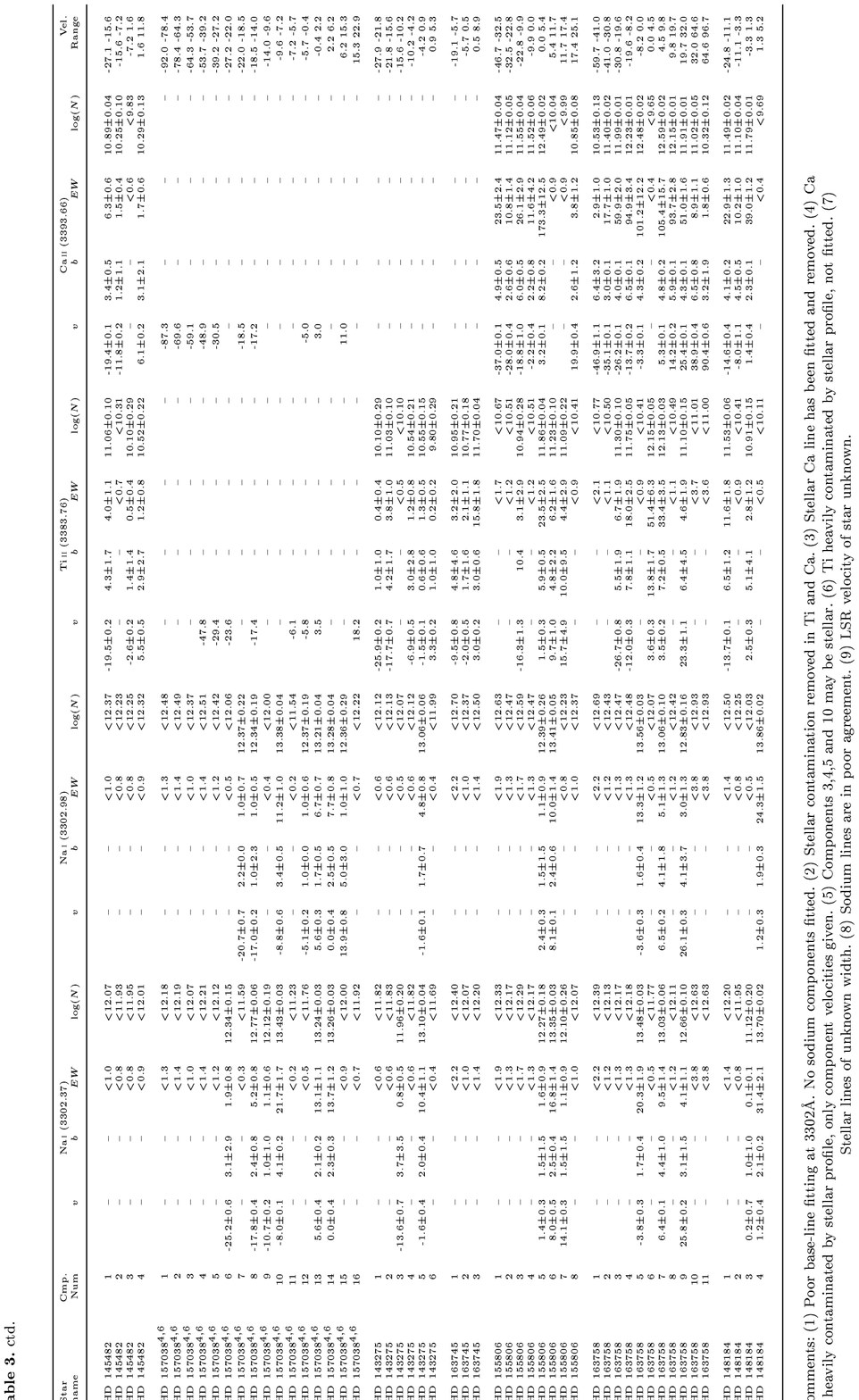, height=245mm,angle=180}
\end{figure*}

\clearpage
\newpage
\begin{figure*}
\epsfig{file=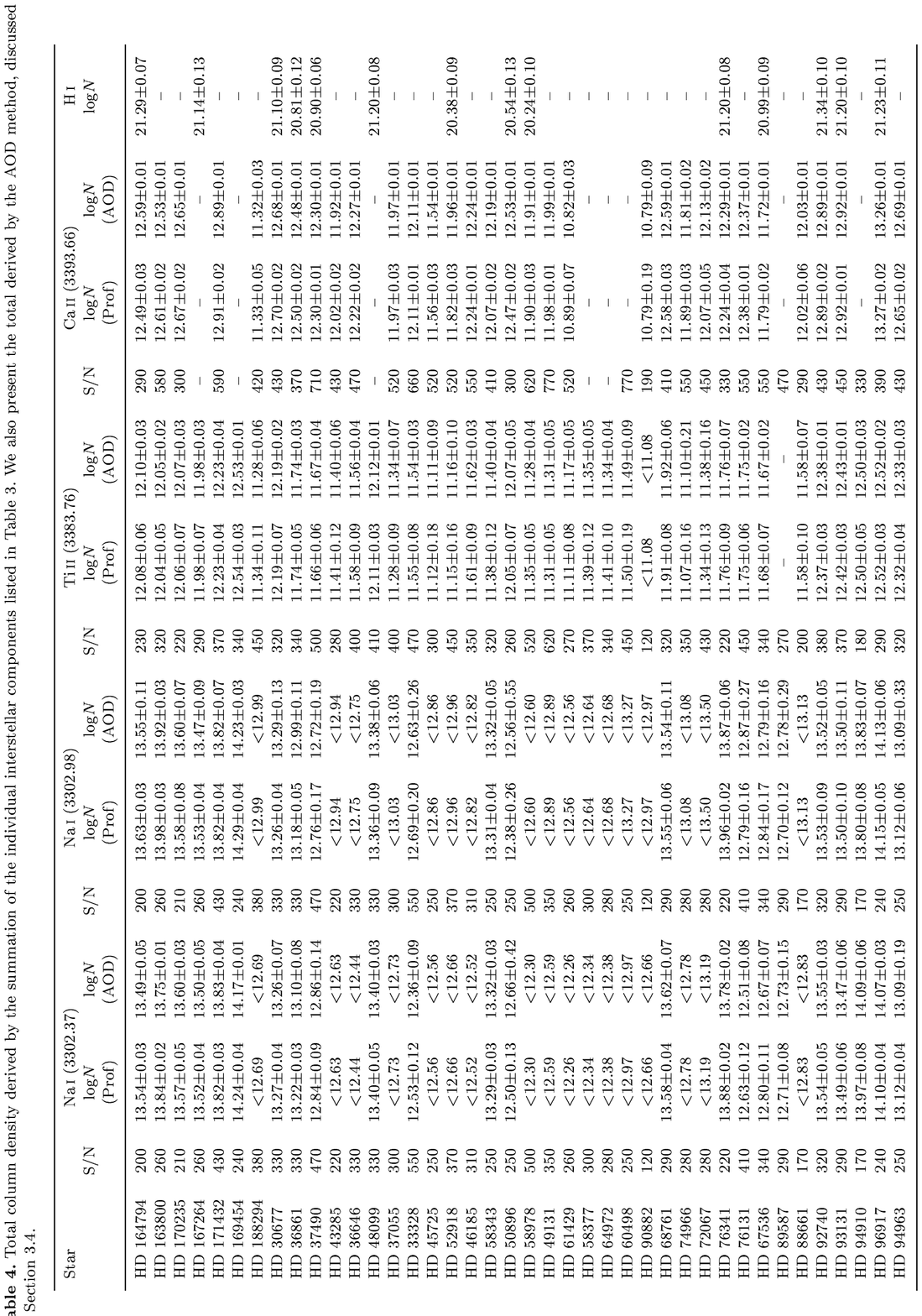, height=245mm, angle=180}
\label{t_totals}
\end{figure*}

\clearpage
\newpage

\begin{figure*}
\epsfig{file=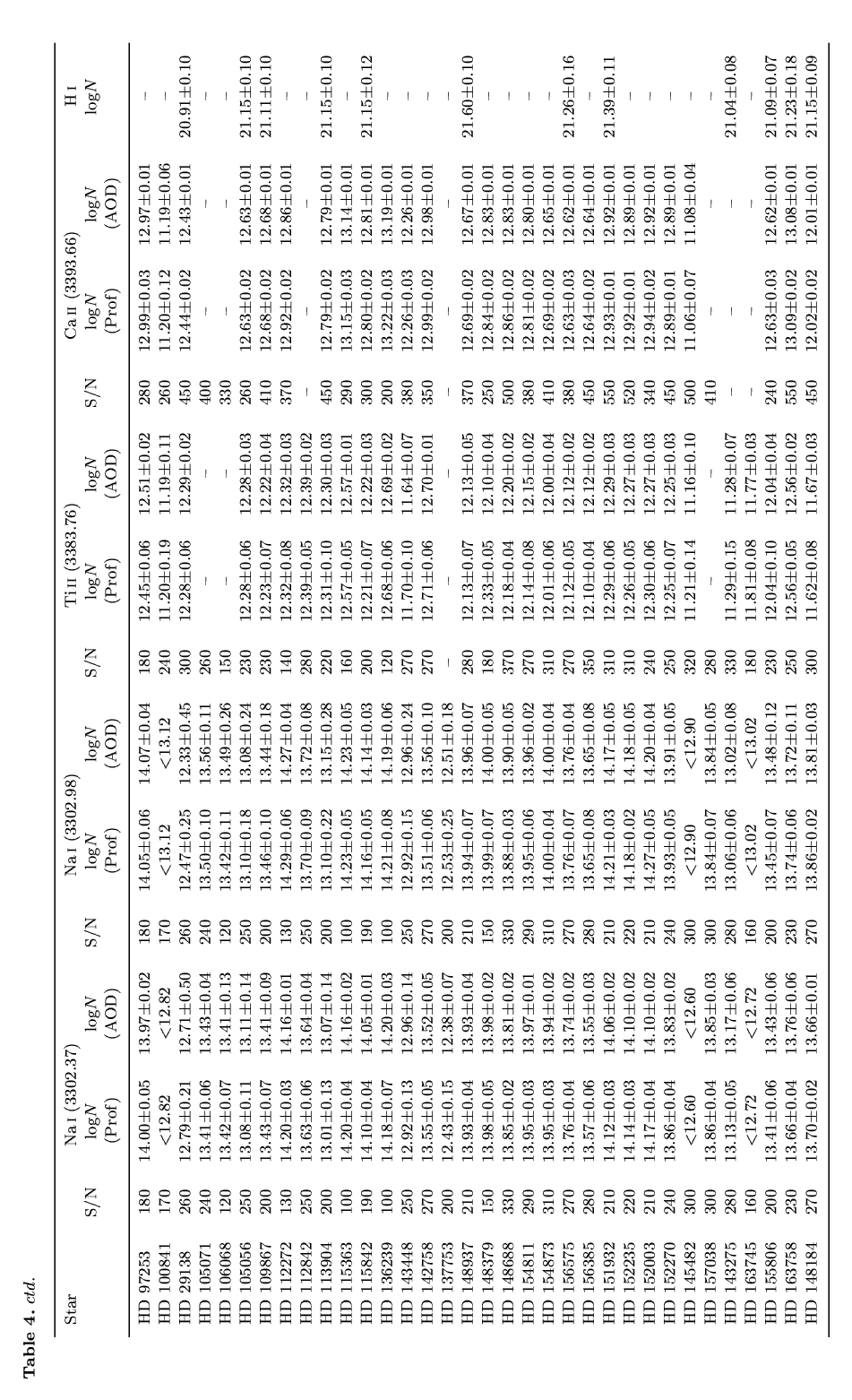, height=245mm, angle=180}
\end{figure*}

\clearpage
\newpage

\setcounter{figure}{3}
\begin{figure*}
\epsfig{file=compareAOD.eps, height=180mm, angle=-90}
\caption[]{The logarithm of the total column density derived from the component
fitting (listed in Table~3) compared to the logarithm of the total column density calculated
using the AOD method. The solid and dashed lines represent
the best-fitting line and the 1$\sigma$ standard deviation in
the points respectively. The RMS scatter and Pearson
correlation coefficients are 0.059, 0.993; 0.055, 0.994;
0.044, 0.997 and 0.027, 0.997 for (a), (b), (c) and (d)
respectively.}
\label{f_compareAOD}
\end{figure*}

\begin{figure}
\epsfig{file=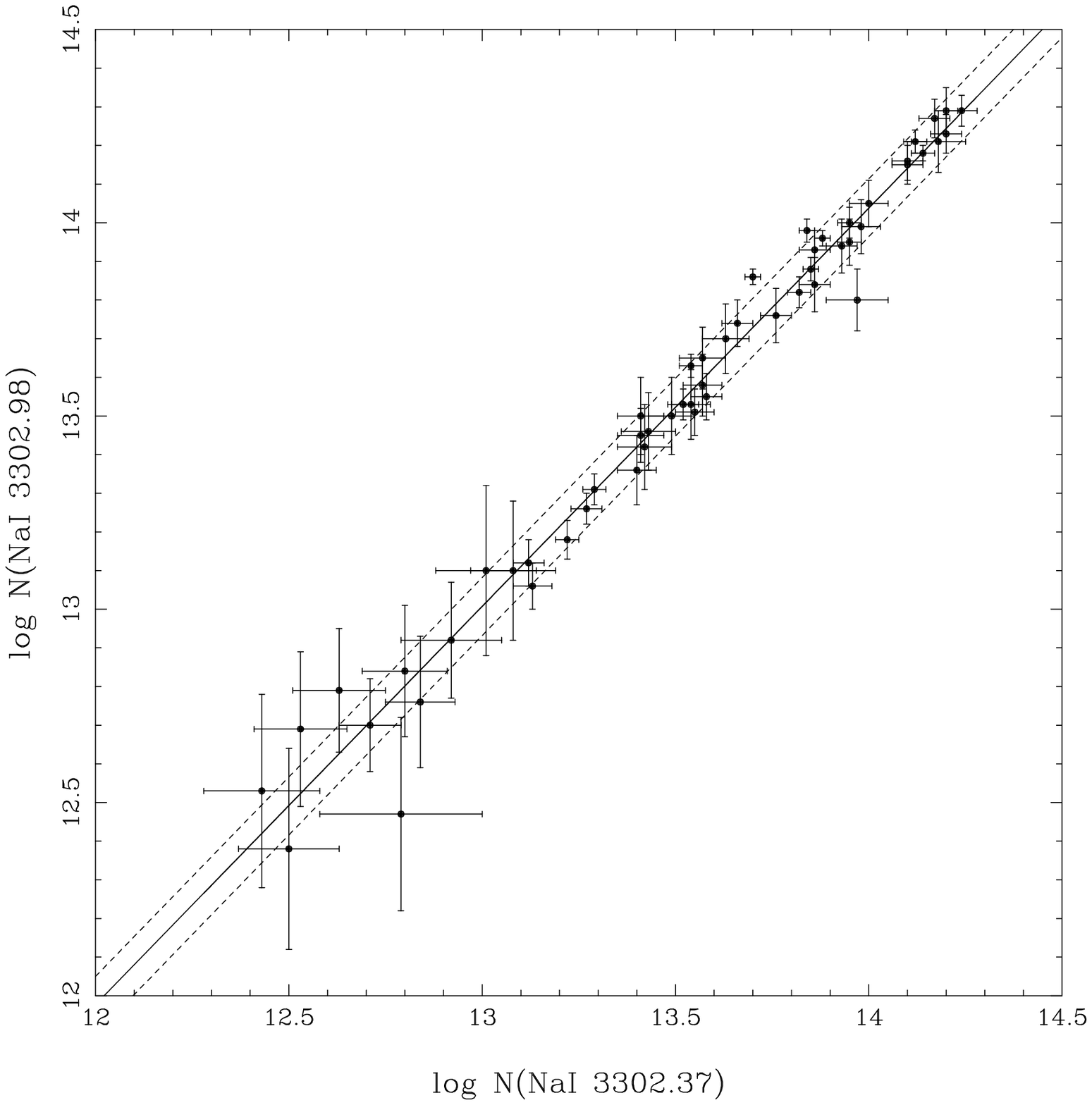, height=90mm, angle=-90}
\caption[]{Comparison of the logarithm of the total column density from each line of the
Na\,{\sc i} UV doublet. The solid and dashed lines represent
the best-fitting line and the 1$\sigma$ standard deviation in
the points respectively. The RMS scatter and Pearson
correlation coefficient are 0.078 
and 0.988 respectively.}
\label{f_totNaIvtotNaI}
\end{figure}

\noindent case, HD\,94910, high velocity features from the weaker line
contaminate the stronger line. 
The 1$\sigma$ standard deviation of the difference in the column density from each line of the Na UV 
doublet is 0.08\,dex, excluding a few discrepant points at low column density the rms is 0.05~dex. 
This value compares with the 3$\sigma$ mode and median error, both of which are 0.06~dex, estimated from the 
component fits as described in Sect.~\ref{s_error_in_com}.

The uncertainty for the Ti and Ca lines should be less than this, as we can be much
more certain about the base-line fit since the S/N ratio is much higher. 
To test this we have calculated the Ca column density from several
Ca\,{\sc ii}~H lines at a range of equivalent width,  
($\lambda_{\rm air}$=3968.468 \AA), and compared these to the column density derived 
from Ca\,{\sc ii}~K.
The average difference in the total is less than 0.01~dex, with the maximum
being 0.02~dex and the lines were close to saturation in this case. Again this
compares well with the total error in the Ca\,{\sc ii} K column density 
estimated in Sect.\ref{s_error_in_com}, which for Ca\,{\sc ii} is 0.02 dex for both the median 
and the mode for the total estimated error. We are thus 
confident that the derived errors in the column densities give a fair approximation of the 
true errors in the analysis. Since the Ti lines have similar structure to the Ca lines, an
average S/N = 300 and do not appear to be saturated, we would not
expect the uncertainty in the total column density to exceed 0.05~dex which is 
the mode of the errors estimated from the component fits as 
described in Sect.~\ref{s_error_in_com}.

In the rest of this paper we take the total Na\,{\sc i} column density as
the average of the values derived from each line. We do not take into
account saturation effects, as in general they are of a similar order to the
standard deviation of the difference in the column density from each line. Because 
the Ca\,{\sc ii} H line is severely blended with stellar H\,{\sc i} and He\,{\sc i} 
features, we do not use this species other than for the above comparison, and our 
Ca column densities are derived from Ca\,{\sc ii} K only.

\subsection{Comparisons with previous work}                   \label{s_comparisons}

From an examination of existing data we have found that four of our stars 
have been previously studied, and column densities determined for some of the
species of interest. In Table~\ref{t_comparisons} we compare our results for the
total column density with those derived in other studies. The agreement between
our Ca\,{\sc ii} total column density and those of other authors is remarkably
good. In the three cases where comparisons could be made, all our values
are within $\pm$\,0.02~dex of the results reported by the other authors. This is to be
expected as the strength of the Ca\,{\sc ii} spectral line is sufficiently
high that it will not be significantly affected by noise in the lower S/N studies of the
other authors. Conversely, the Na\,{\sc i} UV doublet and the Ti\,{\sc ii} are much
weaker and hence we do not expect to obtain as good an agreement, as many of the
weaker features will be obscured by the noise. Indeed, we find
differences of up to 0.33~dex for these lines. Although we have been able to
compare relatively few sightlines, the agreement we have found is encouraging
and we believe that our high S/N spectra allow us to place the best
estimates to date on the column densities of Na\,{\sc i} UV, Ti\,{\sc ii} and
Ca\,{\sc ii} towards each of our stellar targets.

\begin{table}
\setcounter{table}{4}
\caption[]{Comparisons with previous work}
\label{t_comparisons}
\begin{tabular}{llcrc} \hline
Star & Species & \multicolumn{2}{c}{Total Column Density} & Ref.\\
     &         & This Paper & Other Work &\\
\hline
HD\,164794 & Na\,{\sc i}  & 13.59 &    13.58~~~~~ & 1\\
HD\,164794 & Ti\,{\sc ii} & 12.08 &    11.84~~~~~ & 1\\
HD\,164794 & Ca\,{\sc ii} & 12.62 &    12.63~~~~~ & 1\\
HD\,36862  & Na\,{\sc i}  & 13.20 & $>$12.89~~~~~ & 4\\
HD\,36862  & Ti\,{\sc ii} & 11.74 &    11.73~~~~~ & 2\\
HD\,36862  & Ca\,{\sc ii} & 12.50 &    12.52~~~~~ & 4\\
HD\,50896  & Ti\,{\sc ii} & 12.05 &    12.18~~~~~ & 3\\
HD\,50896  & Ca\,{\sc ii} & 12.47 &    12.45~~~~~ & 3\\
HD\,143275 & Na\,{\sc i}  & 13.10 & $>$12.79~~~~~ & 4\\
HD\,143275 & Ti\,{\sc ii} & 11.29 &    11.02~~~~~ & 2\\
\hline
\end{tabular}
\begin{itemize}
\item[] (1)~Welsh et al. \cite{wel97} (2)~Stokes \cite{sto78} (3)~Hobbs
\cite{hob84}  (4)~Hobbs \cite{hob74}
\end{itemize} 
\end{table}


\section{Results}

\subsection{Interstellar clouds along each line of sight}             \label{s_com}

The values of $v$, $b$, $EW$  and log~$N$ are tabulated in Table~3 for each 
component of each
interstellar line observed towards each sightline. We have
associated components in each species based primarily on their velocity. As a 
general rule, components which lie within 3.0~km~s$^{-1}$ have been associated. 
We note that there remains the possibility that even at a resolution of 
3.75 km\,s$^{-1}$, the features detected may be blends of several physically 
different parcels of gas (Welty, Morton \& Hobbs 1996), especially when 
Na is compared with either Ti or Ca, where components are often found not to 
match 
in velocity space. In general, we can usually associate components that 
appear in Ti with 
components in Ca. 

\subsection{Total Column Densities along each line of sight}       \label{s_totals}

In Table~4 we present the total column density for the Na\,{\sc i} UV 
doublet (at 3302\AA), Ti\,{\sc ii} at 3384\AA\ and Ca\,{\sc ii}~K
at 3934\AA. These totals have been calculated as the sum of
the individual components, tabulated in Table~3, along each sightline. 
We also present the S/N ratio measured in the vicinity of
each of the lines, along with the total H\,{\sc i}
column density towards each line-of-sight. The H\,{\sc i} data have been taken from 
the IUE observations of Diplas \& Savage \cite{dip94} for all sightlines, except for 
HD\,50896, HD\,52918 and HD\,148184, where the results 
are from Fruscione et al.
\cite{fru94}, and HD\,58978 from Savage, Edgar \& Diplas
\cite{sav90}.

\subsection{Correlations in total column densities}         \label{s_corr_total}

In this section we discuss correlations in the derived column densities against
parameters such as distance, reddening and H\,{\sc i} column density. Where
best-fits are performed, all slopes assume all free paramaters.

\subsubsection{Correlations with distance}                \label{s_corr_in_dist}

\begin{figure*}
\epsfig{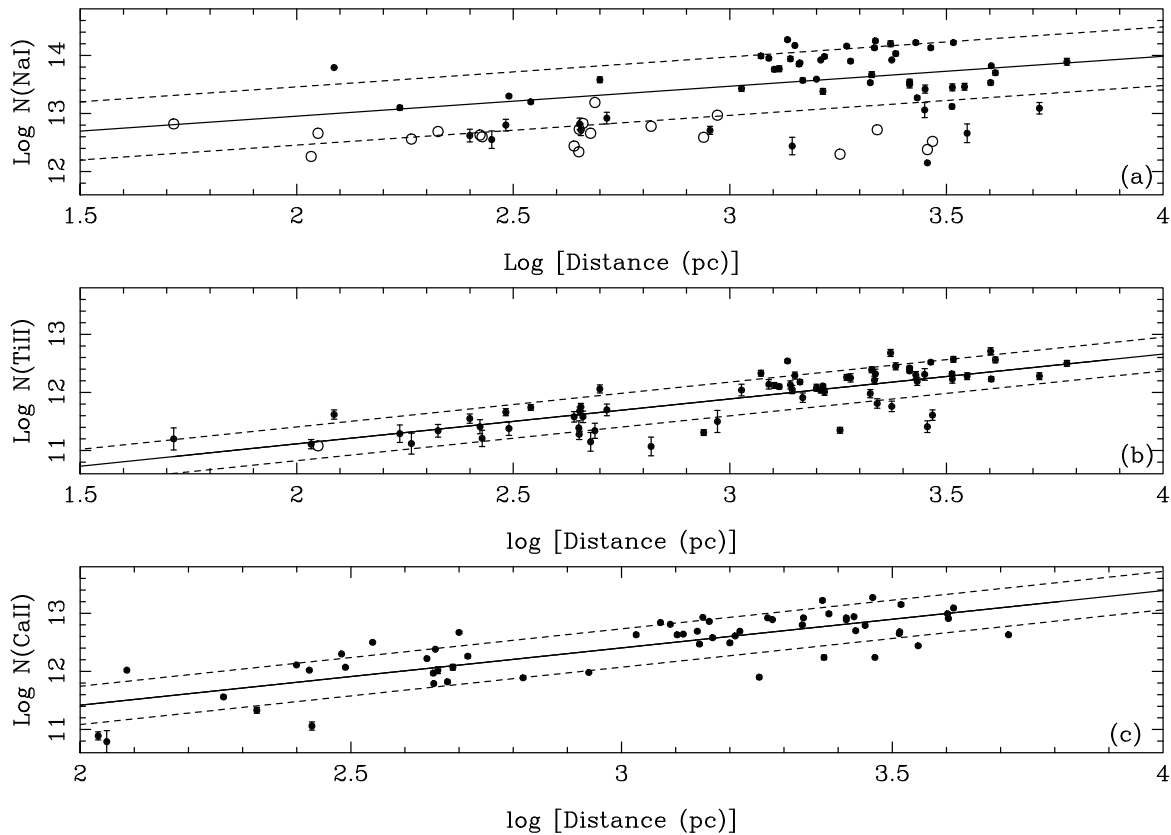}
\caption[]{Plots of the logarithm of the total column density of (a) Na\,{\sc i}, (b)
Ti\,{\sc ii} and (c) Ca\,{\sc ii} against the logarithm of the distance to the star (in pc). Open circles represent
upper limits to the column density.
Open circles are upper limits to the column density. The solid and dashed lines represent
the best-fitting line and the 1$\sigma$ standard deviation in
the points respectively. The error in the distance is 25 per
cent (approximately 0.12 dex in the logarithm). The RMS
scatter and Pearson 
correlation coefficients are 0.501, 0.399; 0.287, 0.776 and
0.330, 0.817 for (a), (b) and (c) respectively.}
\label{f_corr_in_dist}
\end{figure*}

In Fig.~\ref{f_corr_in_dist} we plot the logarithm of the total column density derived from 
profile fitting for each species against the logarithm of the stellar distance (in pc), along
with an un-weighted linear best fit to the data points. 
Fig.~\ref{f_corr_in_dist}(a) shows a weak correlation between the Na
column density and distance, although with the wide scatter of points and small correlation coeffieient 
this correlation may not be significant. However in
Fig.~\ref{f_corr_in_dist}(b) we
see a strong correlation between the Ti column density and the distance,
and a similar strong correlation for Ca in Fig. 6(c), 
both being in agreement with that reported by W97.
This supports other studies which find that Ti and Ca have a 
fairly even spatial distribution, although we should note that there are 
a number of points in which the column density of Ti can vary by as much as one order of
magnitude for stars at similar distances. W97 find this
same spread in the column density and conclude that this is expected due to the
nonuniform distribution of interstellar gas in the spiral arms. In the W97
survey the majority of the stellar targets lie within 1~kpc. However most 
of the targets in our sample lie out to 3.5~kpc. Smoker et al. (2003) observed
a greater depletion of Ca\,{\sc ii} K along sightlines with distances of less
than $\sim$~500~pc and they associate this with the Local Bubble, an elliptical, hot, low-density
region of the ISM which has significant X-ray emissivity 
out to $\sim$~100~pc and is likely to have been formed via a supernova event, see for example, Frisch (1998), 
Frisch \& York (1983). We do not see this effect in  Fig.~\ref{f_corr_in_dist}(c)
but given the limited number of our sightlines which lie within the Local Bubble, 
this is unlikely to be significant.
Finally, we find that the correlations of Ti and Ca with distance are very similar and
this is further evidence that Ti and Ca are mostly associated with the same
interstellar regions.

\subsubsection{Correlations with distance above the Galactic plane}     
                                                             \label{s_corr_in_z}

\begin{figure*}
\epsfig{file=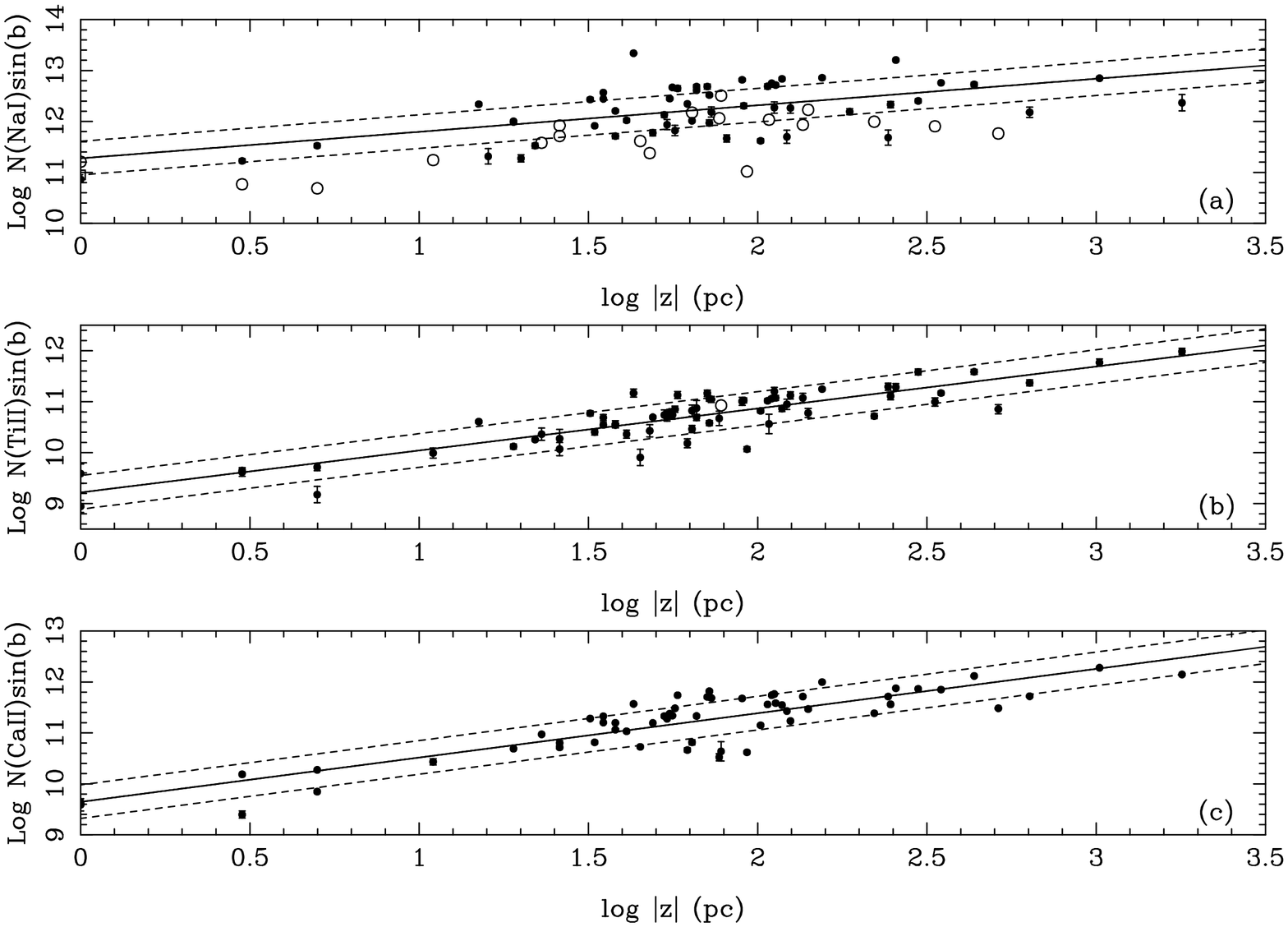, height=180mm, angle=-90}
\caption[]{Plots of the logarithm of the total column density of (a) Na\,{\sc i}, (b)
Ti\,{\sc ii} and (c) Ca\,{\sc ii} against the perpendicular distance to the
star, z, (in pc). Open circles represent
upper limits to the column density.
The solid and dashed lines represent
the best-fitting line and the 1$\sigma$ standard deviation in
the points respectively. The error in the distance is 25 per
cent (approximately 0.12 dex in the logarithm). The RMS
scatter and Pearson 
correlation coefficients are 0.424, 0.564; 0.272, 0.880 and
0.321, 0.864 for (a), (b) and (c) respectively.}
\label{f_corr_in_z}
\end{figure*}

We also examine the correlation between the total column density of each species
and the perpendicular height, $z$, of the stellar target from the
Galactic plane. Following similar methods to W97, we plot
the projected column density, $N_{\rm tot} \times$ sin $b$ against 
$|z|$, along with the 
unweighted linear best-fittings to the
data points, for Na, Ti and Ca in Figs.~\ref{f_corr_in_z}(a), (b) and
(c), respectively. Again we find a weaker correlation in the Na plot, but good
correlations for both Ti and Ca, with the best-fitting lines 
having very similar correlation coefficients of 0.880 and 0.864,
respectively. We note that distance uncertainties contribute an error of
$\sim$0.12 dex, compared with the RMS scatter of 0.29 and 0.33 dex for
Ti and Ca, hence the Ti and Ca relationships 
are remarkably tight. Also, although we  
sample similar Galactic heights to W97, we find a much
tighter distribution of the points about the best-fitting line for Ti and Ca. Hence
we similarly conclude that the column density of each species increases with height
above the Galactic plane, and this trend is in agreement with other sources
such as Edgar \& Savage \cite{edg89}. Although our sample does not
include lines-of-sight with sufficiently high $z$ to calculate the Galactic scale
height of Ti\,{\sc ii} or Ca\,{\sc ii}, our results are in agreement with the 
limits reported. Edgar \& Savage \cite{edg89} also report the smooth distribution
of Ti\,{\sc ii} and Ca\,{\sc ii}, in agreement with our results and they
find that these species are generally associated with the smooth distribution of the
intercloud medium. 

\subsubsection{Correlations with reddening}                \label{s_corr_in_EBV}

\begin{figure*}
\epsfig{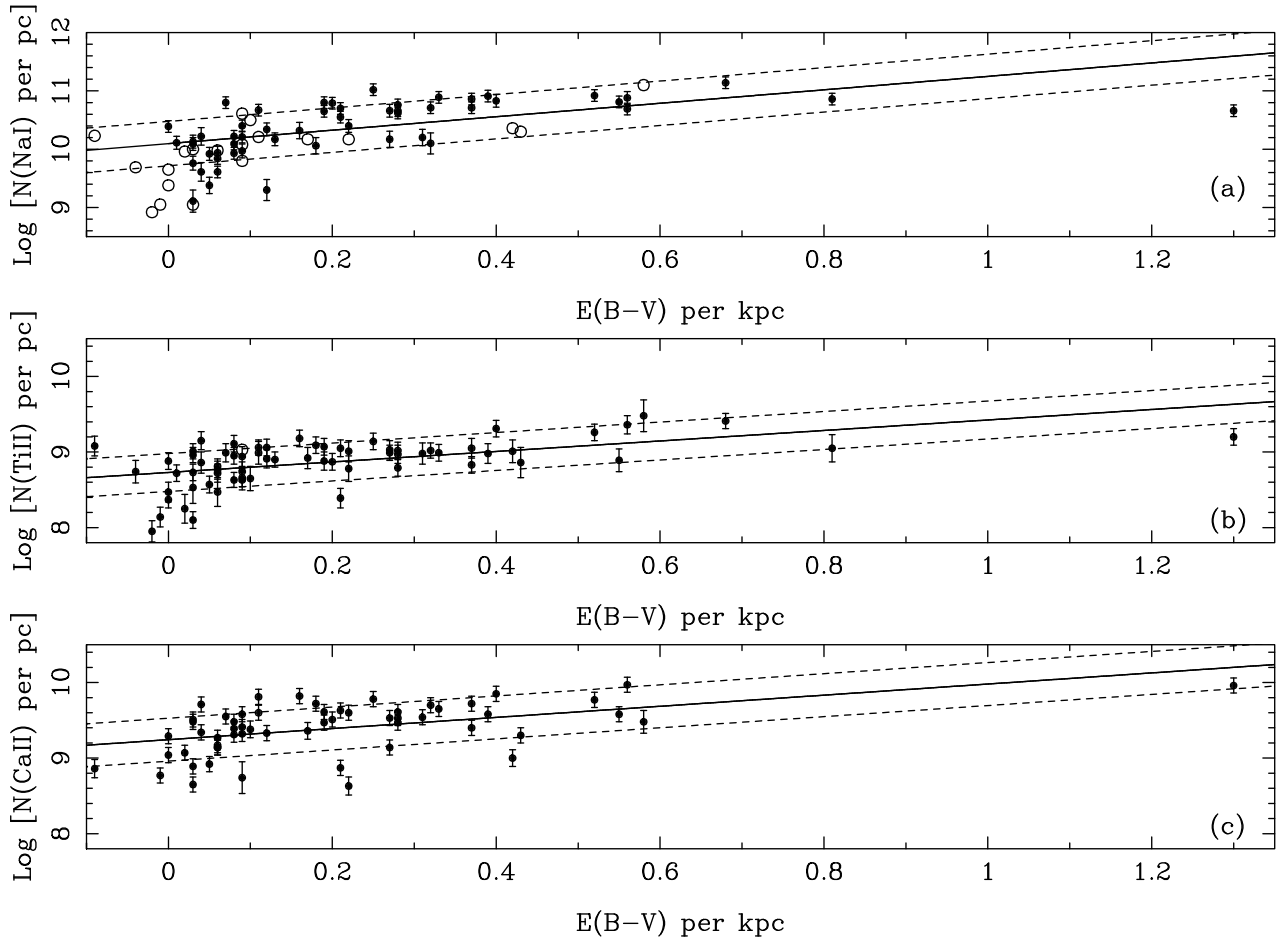}
\caption[]{Plots of the logarithm of the total column density of (a) Na\,{\sc i}, (b)
Ti\,{\sc ii} and (c) Ca\,{\sc ii} per parsec against reddening per kiloparsec. Open circles represent
upper limits to the column density. The uncertainty in the reddening per kiloparsec is approximately
25 per cent.
The solid and dashed lines represent
the best-fitting line and the 1$\sigma$ standard deviation in
the points respectively. The RMS scatter and Pearson 
correlation coefficients are 0.381, 0.597; 0.251, 0.534 and
0.284, 0.486 for (a), (b) and (c) respectively.}
\label{f_corr_in_EBV}
\end{figure*}

In Fig.~\ref{f_corr_in_EBV} we plot the correlation between the logarithm of the total column
density per parsec and the reddening per kiloparsec along the line-of-sight for each of the three species
under investigation. Note that HD\,148184 has been excluded from the plot as it appears 
to have an anomalously high value of redenning per kiloparsec (4.26\,kpc$^{-1}$).

As expected, we find evidence of increasing column density along
sightlines with higher reddenings, as reported by Stokes \cite{sto78} and W97. 
At very low values of reddening, we generally do not observe the Na\,{\sc i} UV doublet
along as many sightlines as we observe Ti or Ca. Hence we see more upper limits
at low reddenings per kiloparsec in Fig.~\ref{f_corr_in_EBV}(a)
compared to Figs. ~\ref{f_corr_in_EBV}(b) and (c). The correlation of the total Na column density 
with reddening appears to be stronger than
that of Ti and Ca. The correlation coefficient in Fig.~\ref{f_corr_in_EBV}(a) is 0.597, compared to 
Figs.~\ref{f_corr_in_EBV}(b) and (c) where the correlation coeffficients are 0.534 and 0.486
respectively. Although the scatter of points in Fig.~\ref{f_corr_in_EBV}(a) is greater than in
Figs.~\ref{f_corr_in_EBV}(b) and (c) this stronger correlation appears
to be significant. This effect is also seen in W97.

\subsubsection{Correlations with total neutral hydrogen column density along the sightlines}                                                      \label{s_corr_in_HI}

\begin{figure*}
\epsfig{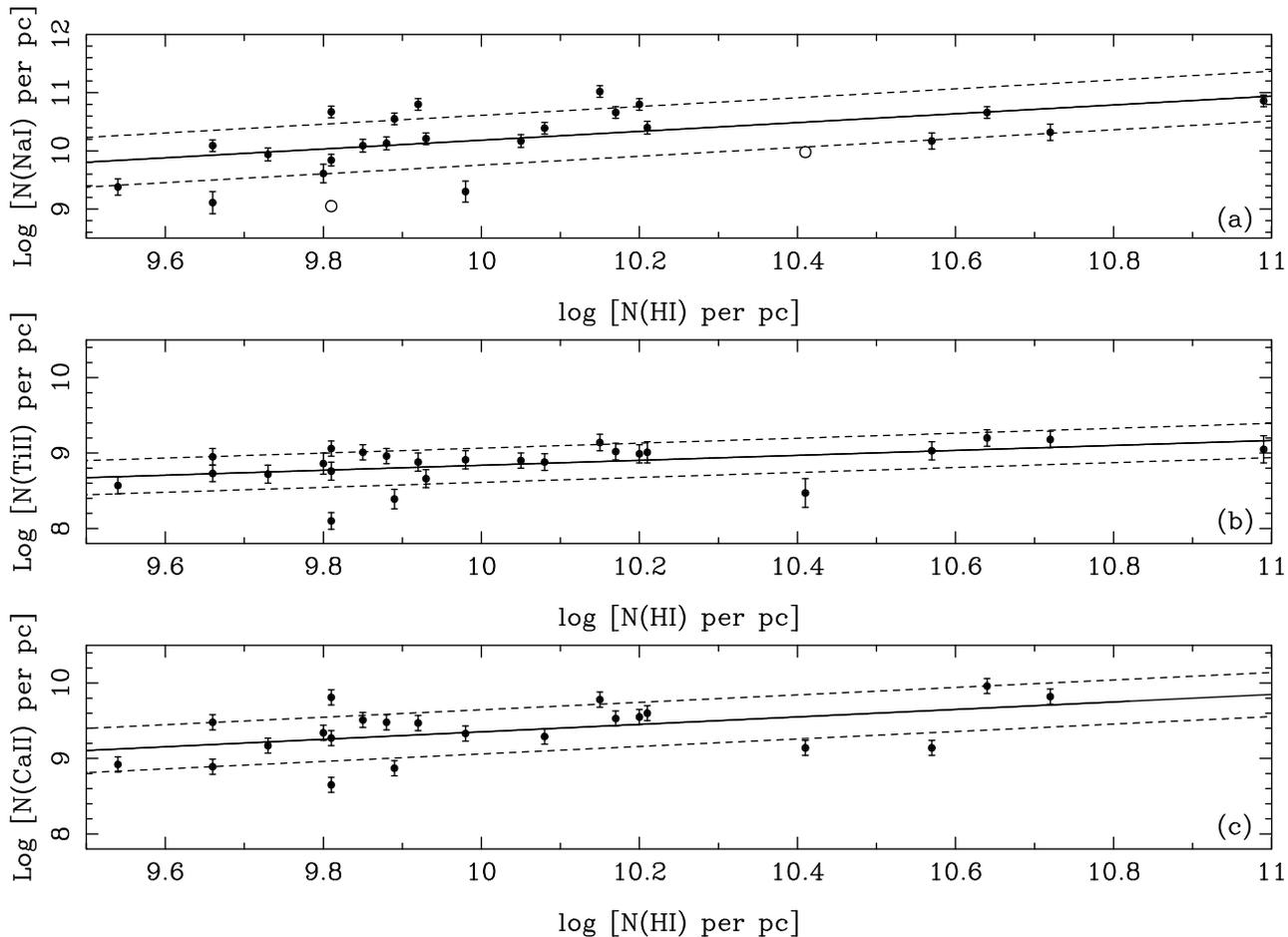}
\caption[]{Plots of the logarithm of the total column density of (a) Na\,{\sc i}, (b)
Ti\,{\sc ii} and (c) Ca\,{\sc ii} per parsec against the logarithm of the total H\,{\sc i} column density per parsec
along the line of sight. Open circles represent
upper limits to the column density. The solid and dashed lines represent
the best-fitting line and the 1$\sigma$ standard deviation in
the points respectively. The uncertainity in the H\,{\sc i}
column density per parsec is typically 0.10~dex. The RMS scatter and
Pearson  
correlation coefficients are 0.427, 0.539; 0.227, 0.459 and
0.293, 0.477 for (a), (b) and (c) respectively.}
\label{f_corr_with_HI}
\end{figure*}

In Figs.~\ref{f_corr_with_HI}(a), (b) and (c) we plot the correlation between
the logarithm of the total column density per parsec and the total neutral hydrogen column density per parsec derived 
from Ly$\alpha$ absorption for each species, taken from Diplas \& Savage \cite{dip94},
Fruscione et al.\cite{fru94} and Savage, Edgar \& Diplas \cite{sav90}. Given the limited number of
sightlines for which we have hydrogen column densities and that we sample a range of
only 1.5~magnitudes in hydrogen column density per parsec it is difficult to derive accurate correlations. From
Fig.~\ref{f_corr_with_HI} it appears that there are similar correlations comparing 
the column density of both Ti and Ca with the
hydrogen column density.
The best-fitting line in Fig.~\ref{f_corr_with_HI}(a) gives a stronger correlation between
the total Na column density per parsec and the total neutral hydrogen column density along the sightline,
although given the larger scatter in Fig.~\ref{f_corr_with_HI}(a) this
may not be significant. The similarity of 
Fig.~\ref{f_corr_with_HI}(b) to Fig.~\ref{f_corr_with_HI}(c) may
indicate that Ti\,{\sc ii} and Ca\,{\sc ii} occur in the same regions of the
interstellar medium. Indeed, we see very similar depletion patterns between Ti\,{\sc ii} 
and Ca\,{\sc ii} and this is fully discussed in Section~\ref{s_depletions}. We note 
that Wakker \& Mathis (2000) also find similar scatters in the 
plots of Ca and Ti column density with H\,{\sc i}
column density (a 1$\sigma$ standard deviation of 0.4 in the log for each species). 
This disagrees with the findings of W97, who 
report a much higher degree of scatter in the Ca plot compared
to Ti, and reasonably conclude that Ca\,{\sc ii} absorption is more affected by
ionization and depletion than is Ti\,{\sc ii}. We note that W97 samples a much larger 
range in $N$(H\,{\sc i}) than do our observations. Our data have values in the rms scatter of 0.29 for 
Ca\,{\sc ii} and 0.23 for Ti\,{\sc ii}. A-priori, the scatter would be expected to be larger in 
Ca\,{\sc ii} than in Ti\,{\sc ii}, as the former is not the primary ionisation stage in the 
warm ISM, although the observed difference is small and may be a statistical variation.

\begin{figure}
\epsfig{file=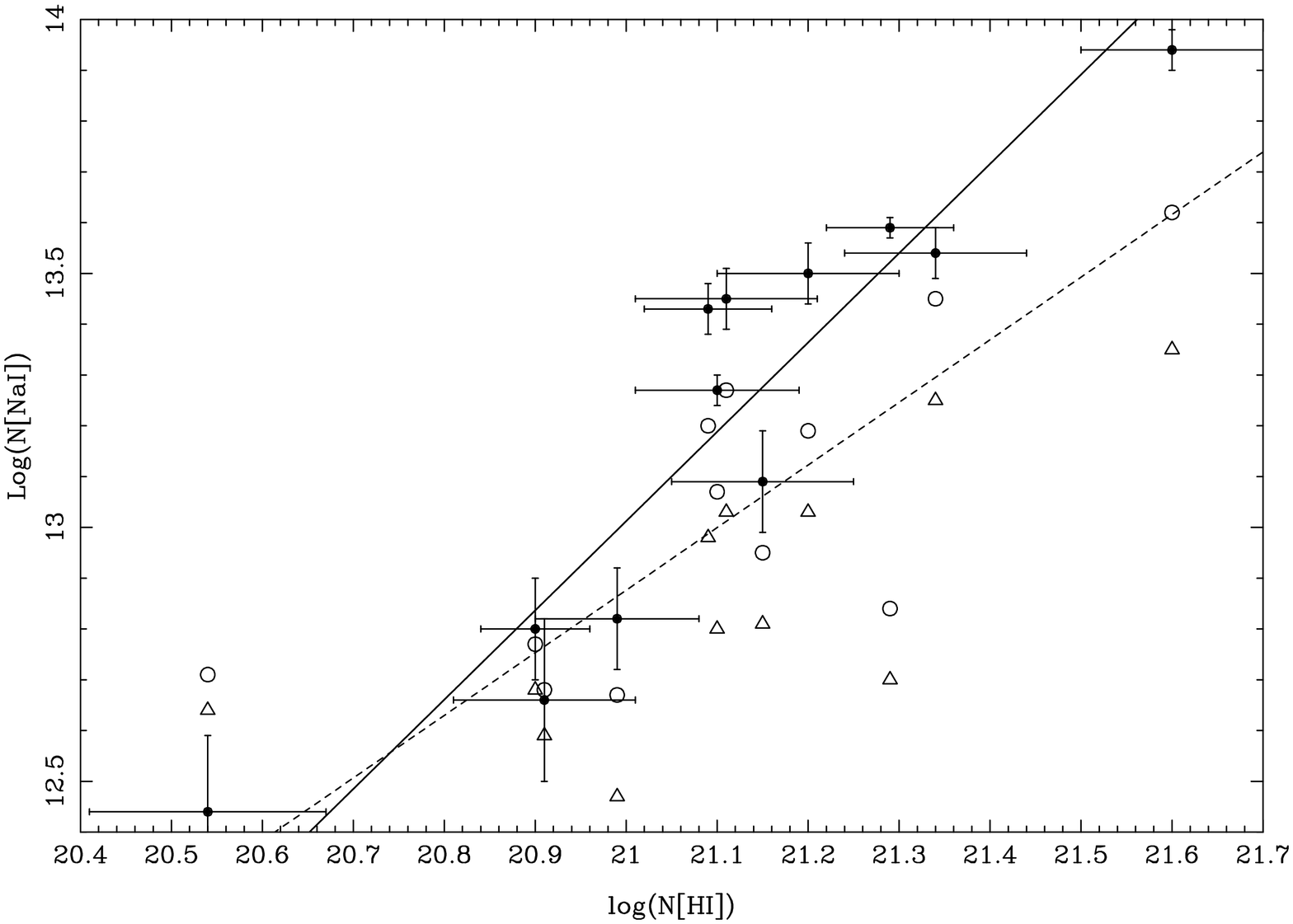, height=90mm, angle=-90}
\caption[]{Correlation between the logarithm of the total Na\,{\sc i} column density derived from a
profile fit of the Na\,{\sc i} UV doublet at 3302\,\AA \, and the logarithm of the total
H\,{\sc i} column density at low reddening, $E(B-V)$~$<$~0.3. The solid line
represents the best-fit to our data points, given as points with error bars. 
The
triangles represent the Na\,{\sc i} column density derived from the weaker line of
the Na\,{\sc i} D doublet at 5893\,\AA. The open circles represent the 
Na\,{\sc i} D
column density corrected for saturation, using the corrections from Savage \&
Sembach \cite{sav91}. The dashed line
represents the best-fitting line to these corrected points. The discrepant point at a H\,{\sc i} column density
of 20.53~dex has not been included in the best-fitting lines.}
\label{f_NaHIrelation}
\end{figure}

Ferlet, Vidal-Madjar \& Gry \cite{fer85} discuss the use of Na\,{\sc i} as a
tracer of H\,{\sc i} in the diffuse interstellar medium via the relationship:

log~$N$(Na\,{\sc i})~=~1.04$\times$log~$N$(H\,{\sc i}~+~H$_{\rm 2}$)--9.09 

\noindent as derived
from Na\,{\sc i}~D lines. These authors
also report that at low reddening ($E(B-V)$~$<$~0.3) this relation is independent
of reddening. In Fig.~\ref{f_NaHIrelation} we plot the total Na column
density against the value for H\,{\sc i} at low reddening, $E(B-V)$~$<$~0.3, and
also show an unweighted best-fitting line to the data points. We note that at 
these reddenings we do not believe the contribution of H$_{\rm 2}$ to be
significant compared to that of H\,{\sc i}. We have also
calculated Na column densities from the weaker line of the 
Na\,{\sc i} D doublet at  5893\,\AA\ for these sightlines where possible, the
results being plotted as triangles. We find excellent agreement between these
results and those predicted by the Ferlet et al. \cite{fer85} relation. From
the best-fitting line to the points plotted as triangles in Fig.~\ref{f_NaHIrelation} (note, this 
best-fitting line is not shown in the figure)
we calculate the relationship between Na\,{\sc i} and H\,{\sc i}
to be:

log~$N$(Na\,{\sc i})~=~1.07$\times$log~$N$(H\,{\sc i})$-$9.66

\noindent This contrasts with the relationship:

log~$N$(Na\,{\sc i})~=~1.76$\times$log~$N$(H\,{\sc i})$-$23.91

\noindent derived from the best fitting line to the Na\,{\sc i} UV points 
shown as the solid line in Fig.~\ref{f_NaHIrelation} although given the scatter
and low number of data-points the significance of this relation is uncertain. A lack
of H\,{\sc i} column densities derived in a consistent manner percludes us from deriving a more
robust relation.

In W97, it is reported that the
ultraviolet Na\,{\sc i} UV doublet can underestimate the true Na\,{\sc i} column
density by up to 20\%. We believe this to be due to the fact that higher velocity
clouds which appear in the Na\,{\sc i}~D lines and are significantly offset from
the strongest feature do not appear in the UV Na\,{\sc i} transitions
due to the inherent weakness of these lines. On the other hand, the high strength of the
Na\,{\sc i}~D lines means that saturation effects cause an underestimate of the
column density for the strongest feature of these lines. This effect can be
much more important than the under-estimation from the UV lines, which goes some way to
explaining the offset between our ultraviolet sodium column density and the column
density from the Na\,{\sc i} D lines. Savage \& Sembach \cite{sav91} estimate the
correction that one needs to apply to account for saturation, based on the
difference between the column density derived from each line of the
doublet. We have applied these corrections to the Na\,{\sc i} D column density and
these are plotted as open circles in Fig.~\ref{f_NaHIrelation}. A best-fitting line to these points gives a relationship of:

log~$N$(Na\,{\sc i})~=~1.23$\times$log~$N$(H\,{\sc i})$-$13.01

\noindent Although this
correction reduces the discrepancy between the column densities from the
Na\,{\sc i} UV and D lines, there is still a significant offset between them. This offset approximately
corresponds to an underestimate of 30 per cent in the column density derived from the
Na\,{\sc i}~D lines compared to using the Na\,{\sc i} UV transitions.

\subsubsection{Correlations between the column density of each species}
                                                \label{s_corr_with_each_species}

\begin{figure*}
\epsfig{file=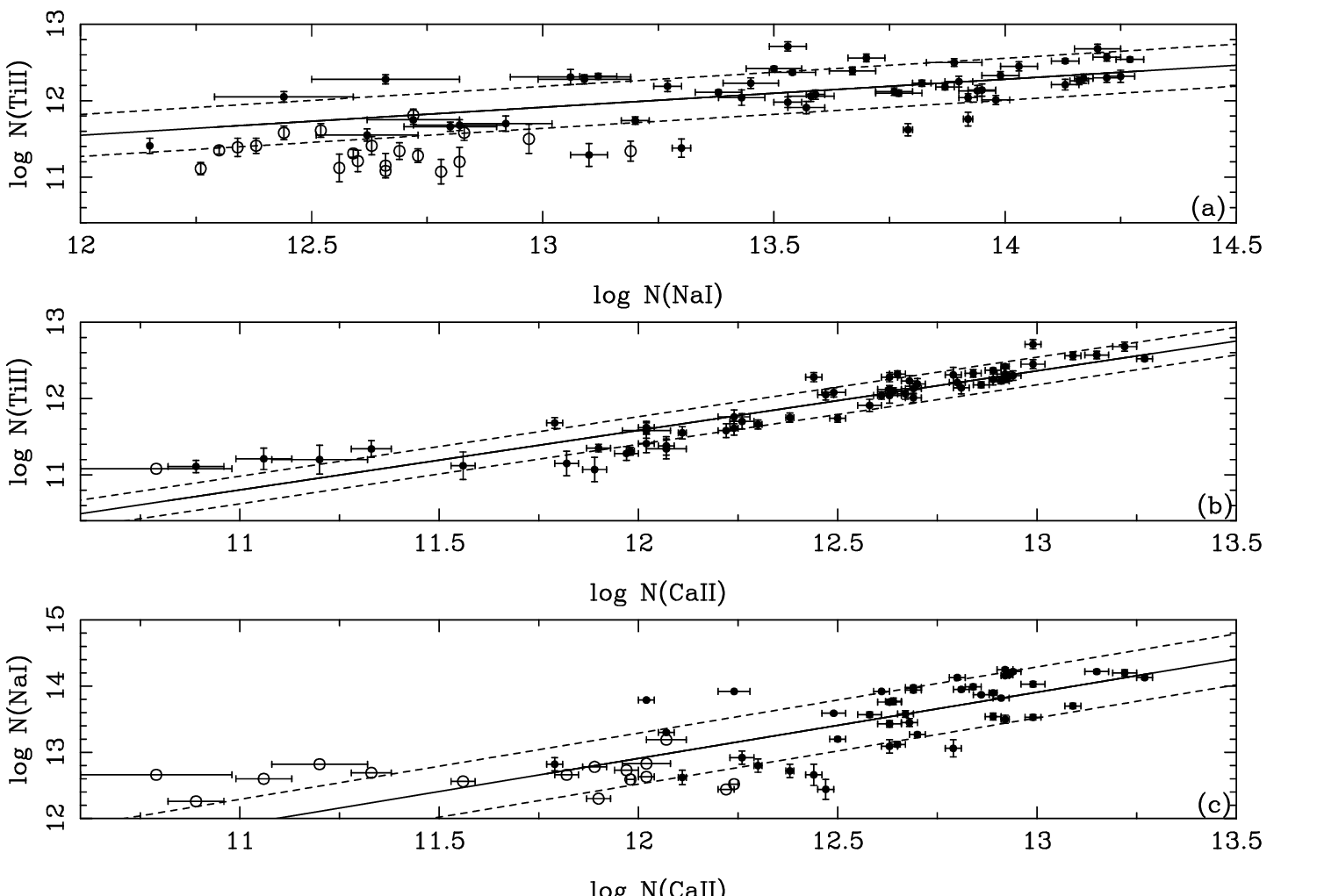, height=180mm, angle=-90}
\caption[]{Correlations between the logarithm of the total column densities derived from each
species. 
The open circles represent upper limits to column densities. The solid and dashed lines represent
the best-fitting line and the 1$\sigma$ standard deviation in
the points respectively. The RMS scatter and Pearson
correlation coefficients are 0.275, 0.573; 0.178, 
0.920, and 0.385, 0.643 for (a), (b) and (c) respectively.}
\label{f_corr_with_each_other}
\end{figure*}

In Figs.~\ref{f_corr_with_each_other}(a), (b) and (c) we present plots of the
correlations between the column densities derived from each species, and we see
good correlation in all three plots. We calculate abundance ratios between the
three species to be 0.04\,$\pm$\,0.01, 0.30\,$\pm$\,0.04 and 0.13\,$\pm$\,0.02
for $N$(Ti\,{\sc ii})/$N$(Na\,{\sc i}), $N$(Ti\,{\sc ii})/$N$(Ca\,{\sc ii}) and
$N$(Ca\,{\sc ii})/$N$(Na\,{\sc i}), respectively, where the quoted uncertainty is
the 1$\sigma$ standard deviation divided by the square root of the number of points
used to determine the ratio. In
Fig.~\ref{f_corr_with_each_other}(b) we see a particularly strong correlation
between the Ti\,{\sc ii} and the Ca\,{\sc ii} abundance and our $N$(Ti\,{\sc
ii})/$N$(Ca\,{\sc ii}) ratio agrees well with the value of 0.29\,$\pm$\,0.03
determined by W97. 

\begin{figure}
\epsfig{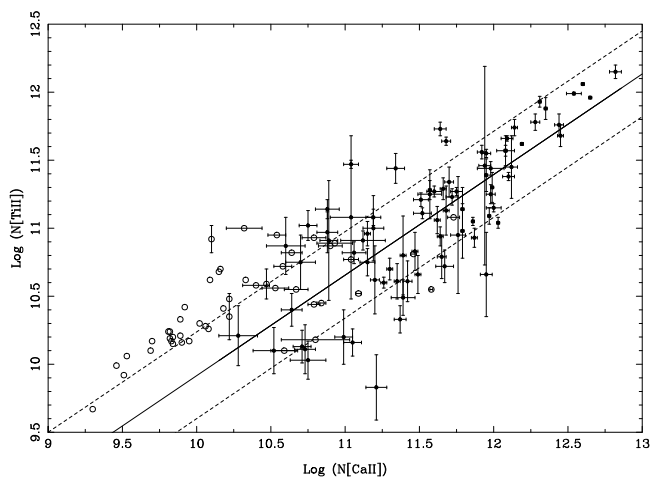}
\caption[]{Correlations between the logarithm of the column density of associated components of
Ti\,{\sc ii} and Ca\,{\sc ii}. The solid and dashed lines represent the best-fitting
line and the 1$\sigma$ standard deviation in the points about this line. The RMS scatter and Pearson 
correlation coefficient are 0.319 and 0.786 respectively.
Open circles represent upper limits to the column density of an individual component.}
\label{f_TiCacomponents}
\end{figure}

In Fig.~\ref{f_TiCacomponents} we plot the column density of individual
Ti\,{\sc ii} components against that 
of the associated Ca\,{\sc ii}
component listed in Table~3. We have not plotted cases where we see two resolved
features in one species and only one feature at a similar velocity in the other
species. Similarly, if there are overlapping components of either Ti\,{\sc ii} 
or Ca\,{\sc ii} that lie within the FWHM of either the main or the overlapping 
component, these components are excluded from the plot. The same rejection 
criteria were applied with respect to the upper limits. The $N$(Ti\,{\sc ii})/$N$(Ca\,{\sc ii}) 
ratio based on the components is 0.26 which is in excellent agreement with that derived from the totals. It
should be noted that in the derivation of this ratio we have not included those
points plotted as open circles. We note that the rms scatter and correlation coeficient 
are larger and smaller respectively than when comparing the total column density along
a line of sight (0.32 dex compared to 0.18 dex for the rms scatter and 
0.78 compared to 0.92 for the Pearson correlation corefficient). This may 
be due to the fact that the components are associated if they lie within 
3 km\,s$^{-1}$ in velocity space, and could infact be separate parcels of 
gas.

Finally, we note that for low upper-limit values 
of Ca\,{\sc ii}, the points appear to be mainly associated with higher 
upper limits for Ti\,{\sc ii}. This may simply be an effect of the higher upper limits for Ti compared to Ca 
upper limits being due to the weakness of the Ti line compared to the Ca line. For the
very low total column densities of Ca, we see a similar effect in
Fig.~\ref{f_corr_with_each_other}(b) although given that this is in only four points on the plots
its significance is in doubt. This is in contrast to W97 (Fig. 8(a)) where the opposite effect is
seen, low column densities of Ca\,{\sc ii} seem to correspond to lower than expected upper limits
to the Ti column denisty. Given that this is only seen in the upper limits and not for any real detections this
may be an effect of the methodology which was used to estimate the upper limits. In 
W97 Fig. 8(c) where they plot Ti against Ca for individual components, this effect 
is not seen.

\subsubsection{Depletions in the column density along each line-of-sight}
                                                            \label{s_depletions}

\begin{figure*}
\epsfig{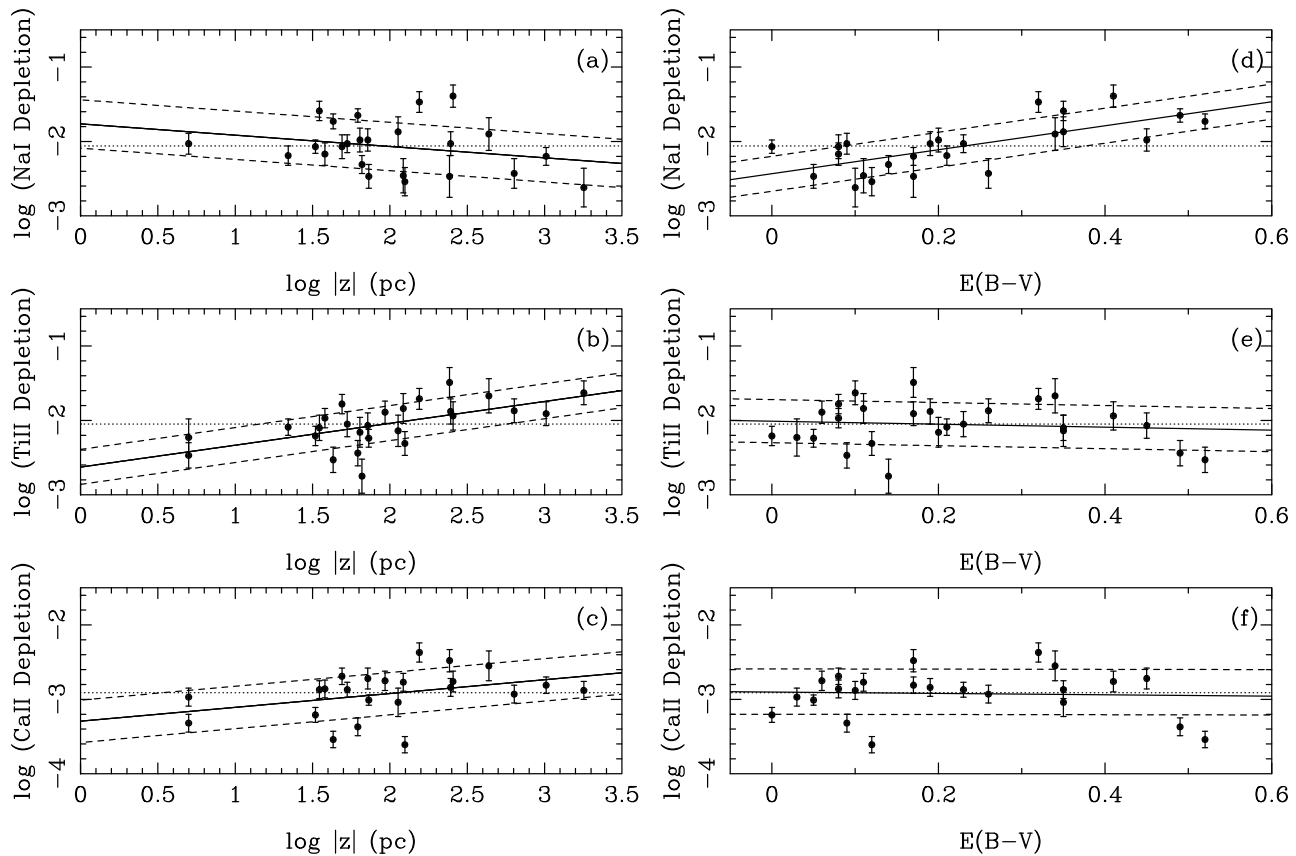}
\caption[]{Panels (a), (b) and (c) show the depletion as a function of the
perpendicular distance to the star to the Galactic plane for Na\,{\sc i}, Ti\,{\sc ii} and
Ca\,{\sc ii} respectively. Panels (d), (e) and (f) show the depletion as a function of
the reddening along the sightline to the star for Na\,{\sc i}, Ti\,{\sc ii} and
Ca\,{\sc ii} respectively. The dotted line represents the average depletion
derived from our sample. Open circles represent lower limits to the depletion. The solid and dashed lines represent
the best-fitting line and the 1$\sigma$ standard deviation in
the points respectively. The RMS scatter and Pearson 
correlation coefficients are 0.322, -0.253; 0.236, 0.592; 
0.285, 0.370; 0.239, 0.695; 0.291, -0.101 and 0.306, -0.046 for
(a), (b), (c), (d), (e) and (f) respectively}.
\label{f_Depletion}
\end{figure*}

In Fig.~\ref{f_Depletion} we plot the depletion of the species with respect to
the perpendicular height and the reddening along the line-of-sight, in a similar
method to that used by W97. The depletion is calculated as
the ratio of the abundance of the element to the solar system value. 
Solar system abundances of 6.37, 5.00, and 6.41 for Na, Ti and Ca, respectively, are
obtained from Lodders \cite{lod03}, where the abundance of the element is taken
as its total column density relative to the H\,{\sc i} column density along the
line-of-sight. Previous work (e.g. Gondhalekar 1985, 
Crinklaw, Federman \& Joseph 1994) indicates that the depletion of both Ti and Ca 
increase in line with increasing column density. Both elements show the same 
slope with increasing column density, indicating that these elements only trace the 
warm intercloud medium. In the current paper we hence restrict ourselves to 
correlations in depletion of the three species studied with distance above the 
Galactic plane and reddening in line with W97. 

Fig.~\ref{f_Depletion}(a) shows a weak correlation between the
height, $|z|$, and the depletion for Na\,{\sc i}, in the 
sense of a higher depletion at high $z$, but given the scatter in the plot (Pearson 
correlation coefficient of -0.25) this 
does not appear significant. In Figs.~\ref{f_Depletion}(b)
and (c) we see evidence for lower depletions of Ti and Ca at higher values of
$|z|$, consistent with less dust with increasing distance from the Galactic plane. 
In Fig.~\ref{f_Depletion}(d) we see a correlation between the depletion 
of Na and the reddening along the line-of-sight, with again less depletion at 
higher reddenings. Given the small scatter and large correlation coefficient, 
the correlation appears to be significant. In contrast, we find evidence of 
constant depletions of Ti and Ca at all reddenings, shown in Figs. 
~\ref{f_Depletion}(e) and (f). This disagrees with 
the work of de Boer et al. (1986) who find that the depletion of all metals 
increases with increasing E(B-V), although for Ti\,{\sc ii} there were only 8 datapoints. 
However, our finding of a lack of correlation agrees with W97 who find little 
evidence in Ti\,II for such an effect with E(B-V) $>$ 0.2. We note that the 
scatter and range in values of depletion found in the current paper for Ti\,{\sc ii} 
of -1.4 to -2.8 dex are somewhat smaller than found by W97 whose depletion 
values range from -0.5 to -2.7 dex. 
 
Considering the decreasing Na depletion with increasing E(B-V), we associate the 
sightlines with higher reddenings as having a significant amount of H$_{2}$. We 
have been able to obtain the H$_{2}$ column density for one line-of-sight, 
HD\,148184, at high reddening (E(B-V)=0.53) from Fruscione et al. \cite{fru94}, 
and we find that this is 30\% of the H\,{\sc i} column density. For an element 
that is primarily in clouds, the exclusion of H$_2$ will cause the total 
observed H column density to be decreased, causing an observed increase in the 
abundance of the element and a corresponding decrease in depletion which is 
what we observe in Fig.~\ref{f_Depletion}(d) at high values of reddening. On 
the other hand, regions with higher E(B-V) values are associated with dust onto 
which elements are removed from the gas-phase, which causes an increased depletion. 
Hence the two affects act in different directions. We note that e.g. Phillips, 
Pettini \& Gondhalekar (1984) find that over a range of column densities, the 
depletion of sodium is essentially constant. 


\subsubsection{$N$(Ti\,{\sc ii})/$N$(Ca\,{\sc ii}) ratio}               \label{s_Ti/Ca}

\begin{figure*}
\epsfig{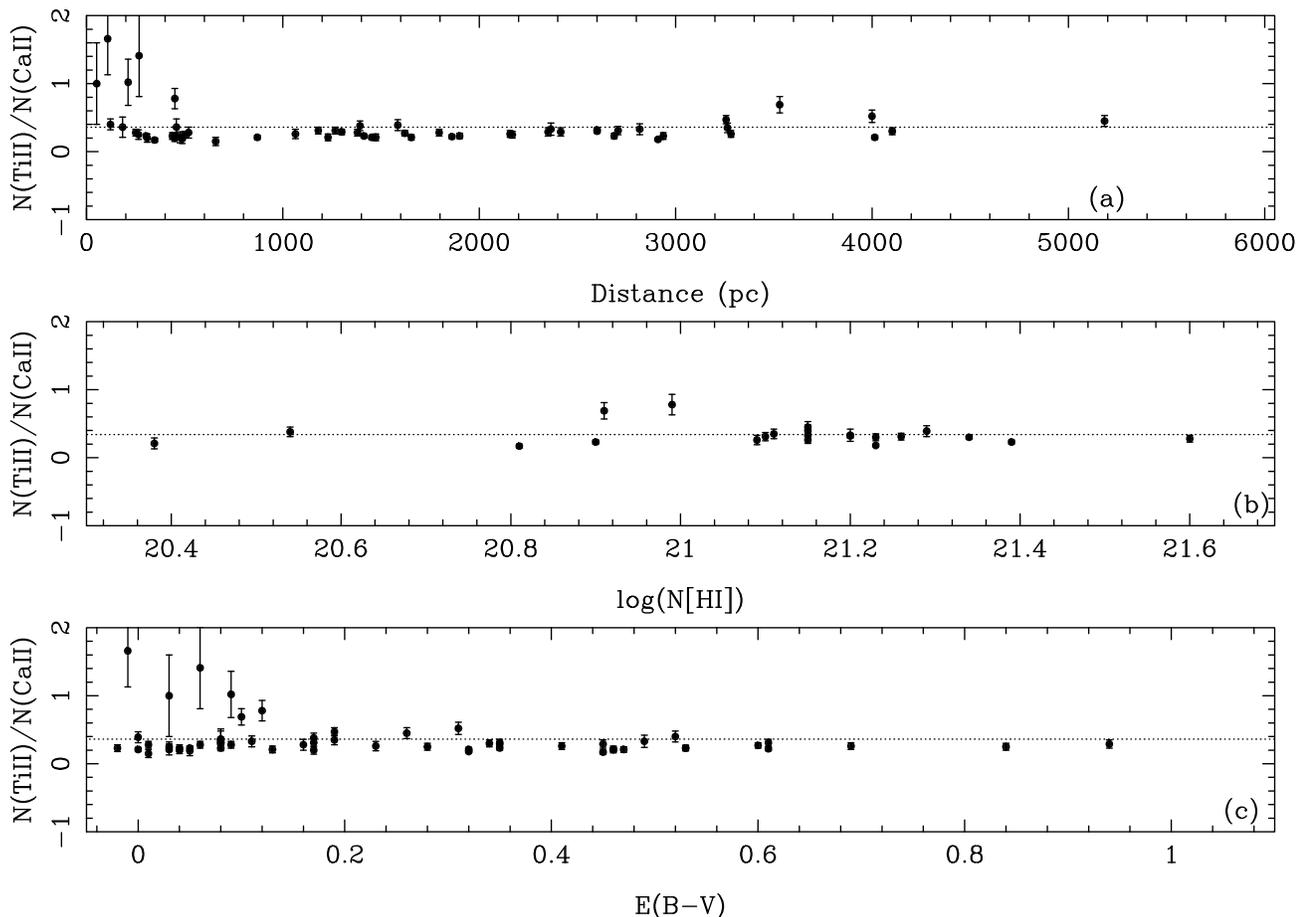}
\caption[]{(a), (b) and (c) plot the $N$(Ti\,{\sc ii})/$N$(Ca\,{\sc ii}
ratio as a function of distance to the target, total H\,{\sc i} column density and
reddening respectively. The dotted lines represent the average $N$(Ti\,{\sc
ii})/$N$(Ca\,{\sc ii}) ratio.} 
\label{f_TiCaratio}
\end{figure*}

In order to reliably conclude that Ti\,{\sc ii} and Ca\,{\sc ii} do indeed occur
in similar environments we plot the $N$(Ti\,{\sc ii})/$N$(Ca\,{\sc ii}) ratio
against distance to the stellar target, total H\,{\sc i}
column density along sightline
and reddening in Fig.~\ref{f_TiCaratio}(a), (b) and (c),
respectively. We also plot the average $N$(Ti\,{\sc ii})/$N$(Ca\,{\sc ii}) ratio as
a dotted line in these figures. Fig.~\ref{f_TiCaratio}(a) shows 
a number of ``discrepant'' points, i.e. with higher than average 
$N$(Ti\,{\sc ii})/$N$(Ca\,{\sc ii}) 
ratios, towards and HD\,100841 and HD\,145482. These 
sightlines have in common that, given errors in their distance estimates, they can be associated within 
the Local Bubble. The only peculiarities in these 
sightlines are that for Ca\,{\sc ii} K in HD\,100841 the lines are broad and weak
and for HD\,145482 the lines are also relatively 
weak. From each of these plots we conclude that, in general, 
$N$(Ti\,{\sc ii})/$N$(Ca\,{\sc ii}) ratio is constant 
over distance, cloud density and reddening, and hence we can infer that 
Ti\,{\sc ii} and Ca\,{\sc ii} normally occur in the same regions. The RMS of the scatter when we 
plot the total column density of Ti\,{\sc ii} against Ca\,{\sc ii} is some 0.18 dex. The mode in the 
errors in the total column densities for each species are 0.02 dex for Ca\,{\sc ii} and 0.05 dex 
for Ti\,{\sc ii}, so the intrinsic scatter on the data is only some 30--40 per cent. We recall that the 
observed depletion is dependent in each sightline on the fraction of warm and cold gas present in each 
component (Spitzer 1985), the dust content (ions being removed from the gas-phase by sticking onto grains) 
and the ionizing radiation field. The fact that Ca\,{\sc ii} does not vary much with respect to 
Ti\,{\sc ii} was initially surprising, as it suggests that the amount
of Ca\,{\sc ii} (not the dominant ionisation stage) does not 
change much with the radiation field. However, at least when 
comparing the warm intertellar medium to low-density H\,{\sc ii} regions, the work of Sembach et al. 
(2000) predicts that the relative fraction in these transitions 
remains constant. For example, in their composite 
model of the warm interstellar medium (WIM) by Sembach et al. (2000), the ionisation fraction is 0.21 
for Ca\,{\sc ii}, 0.79 for Ca\,{\sc iii},  0.59 for Ti\,{\sc ii}, and 0.41 for Ti\,{\sc iii}. For WIM regions 
with the fraction of neutral hydrogen x$_{\rm edge}$, is greater than 0.1, the corresponding values of  
the ionisation fraction for their model is 0.19 for Ca\,{\sc ii}, 0.79 for Ca\,{\sc iii},  
0.45 for Ti\,{\sc ii} and 0.52 for Ti\,{\sc iii}. Finally, for the x$_{\rm edge}$=0.1 model for 
a low-density H\,{\sc ii} region, the ionisation fractions are 0.06 for Ca\,{\sc ii}, 0.94 for Ca\,{\sc iii},  
0.18 for Ti\,{\sc ii} and 0.63 for Ti\,{\sc iii}. In these models at least, the 
relative Ti\,{\sc ii}/Ca\,{\sc ii} ionisation fractions are thus a factor 3 different in both cases. Hence if 
the models are accurate and the sightline is comprised only of WIM and H\,{\sc ii} regions, the 
strong correlation between Ti\,{\sc ii} and Ca\,{\sc ii} is not unexpected. Of course, 
the above discussion does not take into account the presence of dark clouds in which both 
elements are heavily depleted.

\subsection{Components}

\subsubsection{Intermediate and High velocity components}         \label{s_IVC_HVC}

In order to classify the interstellar components we have 
observed as low-, medium-
or high velocity (LVCs, IVCs and HVCs, respectively), we have
calculated the expected velocity range in which one
would expect to see interstellar clouds based on the direction of the sightline
and the distance to the stellar target. To calculate the velocity range we use
the methodology of Wakker \cite{wak91}, in that we assume a flat rotation curve
with $v$$_{\rm rot}$~=~220~km~s$^{-1}$ at $r>$0.5 kpc, decreasing linearly 
towards the Galactic centre, together with equations from Mihalas \&
Binney \cite{mih81}.  A deviation velocity for interstellar cloud
components which lie outside the expected velocity range is calculated, 
where the deviation
velocity is defined as the difference between the velocity of the component and
the nearest limit of the expected velocity range (Wakker 1991). We
classify LVCs as having absolute values of their deviation velocities below 30~km~s$^{-1}$, 
IVCs between 30~km~s$^{-1}$ and 90~km~s$^{-1}$, and HVCs 
greater than 90~km~s$^{-1}$. In Table~\ref{t_IVC_HVC} we
tabulate all the IVCs and HVCs detected along our sightlines. As the present
paper is not primarily concerned with IVCs and HVCs, and we also do not have
access to H\,{\sc i} emission spectra at present, we are hence unable to
confirm that our detected IVCs and HVCs are real interstellar clouds. 
We must emphasise that given the nearby nature of many of the stars, it 
is unlikely that they are true HVCs/IVCs but this possibility should not be discounted. 
Although high velocity clouds are thought to reside either in the Galactic halo or possibly 
at extragalactic distances (e.g. Wakker \& van Woerden 1997, Braun \& Burton 1999), they can also arise from
shocks created in the ISM by supernova blast waves.  
In the present paper we will restrict ourselves to a brief discussion
of several of the more interesting sightlines, although a full study
of both IVCs and HVCs will be carried out in a future paper.

\subsubsection{HD\,171432}

We see 5 IVCs in the Ca\,{\sc ii}~K line and these have also been
identified in the Ca\,{\sc ii}~H feature and hence these lines are unlikely to be stellar transitions
of another species. We have correlated our sightline with data from the
H\,{\sc i} survey of high velocity interstellar clouds in the Southern
Hemisphere (Morras et al. 2000), and we do not find any nearby H\,{\sc i} clouds
at similar velocities to our HVCs, although we have not inspected the actual 
H\,{\sc i} data. 

HD\,171432 lies along the line of sight to the
Scutum Supershell and Callaway et al. \cite{cal00} report a kinematically-derived 
distance to the Supershell of 3.3~kpc with a diameter of $\sim$290-pc. Our spectroscopic 
distance estimate puts the star at a distance of 4.0 kpc and so the IVC's appear to  be
associated with this feature.  
We also see some evidence for IVCs in the Ti\,{\sc ii} spectra although these have not been fitted. On closer
inspection of the spectra, we see evidence of many small strength features which do
not appear to be noise. Based on the FWHM of the stellar lines of the star, these features
are clearly not stellar in nature (unless there is a binary companion present), but they may be
evidence of small scale structure in the Scutum Supershell.

\subsubsection{HD\,72067}                                \label{s_HD72067}

We see a single strong HVC along this sightline which also
appears in Ca\,{\sc ii}~H. This sightline lies along the same 
sightline as the Vela supernova remnant which Cha, Sembach \& Danks \cite{cha99} report 
to lie at a distance of 250~pc. Our distance estimate of 488~pc would place HD\,72067 within and possibly beyond
the Vela supernova remnant and as many previously observed HVC's have been seen along sightlines which pass through this 
remnant, for example, as reported by Slavin, Nichols \& Blair (2004), our observation of an HVC is not unexpected.

\subsubsection{HD\,94910}

This is the only sightline in which we detect an HVC in the Na\,{\sc i} UV 
data, and also appears in the Ca\,{\sc ii}~K line. 
Note that we have not fitted the interstellar components of the Ca\,{\sc
ii}~K line due to the difficultly of accurately fitting a base-line to the
continuum. The {\sc simbad} database classifies HD\,94910 as a Wolf-Rayet star and
the high velocity feature we see may be associated with the star rather than the
interstellar medium, as Wolf-Rayet stars are known to have many peculiar
features
(Leitherer \& Chavarria 1987), such as stellar-wind bubbles this could explain our 
high
velocity features. Indeed, we find that over two thirds of our stellar targets
which are classified as Wolf-Rayet's show IVCs or HVCs.

\subsubsection{HD\,163758}

Again we have confirmed the existence of the high velocity feature in the
Ca\,{\sc ii}~H line, although similarly to HD\,94910 this star is classified as a
Wolf-Rayet star and as such the HVC may be a circumstellar feature. 
A search of the Morras et al. (2000) H\,{\sc i} database does not find any
nearby HVCs at a similar velocity to our detected feature.

\begin{table*}
\begin{center}
\small
\caption{List of sightlines showing intermediate and high velocity features 
in absorption. 
}
\label{t_IVC_HVC}
\begin{tabular}{lrrrcrrrrrrr}
\hline
  Star     & $l$~~ & $b$~~ & $d$(pc)& Cmp.  & \multicolumn{2}{c}{Na\,{\sc i}(3302)}&\multicolumn{2}{c}{Ti\,{\sc ii}(3384)}&\multicolumn{2}{c}{Ca\,{\sc ii}(3934)}& Type       \\
	   & (deg.)& (deg.) &	     & No.   &$v_{\rm LSR}$&$v_{\rm Dev}$&$v_{\rm LSR}$&$v_{\rm Dev}$&$v_{\rm LSR}$&$v_{\rm Dev}$&	  \\
\hline
	   &	   &	   &	    &	     &  	     &  	     &  	     &  	     &  	 &	   &	  \\
  HD 164794&   6.01&  -1.21&    1585&       2&        --  &         -- &       -31.2&       -31.2&        --  &         -- &       IVC  \\
  HD 167264&  10.46&  -1.74&    2109&       5&        --  &         -- &        43.7&        30.9&        --  &         -- &       IVC  \\
  HD 171432&  14.62&  -4.98&    4014&       1&        --  &         -- &        --  &         -- &       -65.2&       -65.2&       IVC  \\
  HD 171432&  14.62&  -4.98&    4014&       2&        --  &         -- &        --  &         -- &       -62.5&       -62.5&       IVC  \\
  HD 171432&  14.62&  -4.98&    4014&       3&        --  &         -- &        --  &         -- &       -47.2&       -47.2&       IVC  \\
  HD 171432&  14.62&  -4.98&    4014&      12&        --  &         -- &        --  &         -- &        92.6&        48.7&       IVC  \\
  HD 171432&  14.62&  -4.98&    4014&      13&        --  &         -- &        --  &         -- &       117.4&        73.5&       IVC  \\
  HD 169454&  17.54&  -0.67&    1359&       1&        --  &         -- &       -40.2&       -40.2&        --  &         -- &       IVC  \\
  HD  37490& 200.73& -14.03&     304&       1&        --  &         -- &        --  &         -- &       -45.7&       -45.7&       IVC  \\
  HD  50896& 234.76& -10.08&    1393&       1&        --  &         -- &        --  &         -- &       -65.0&       -65.0&       IVC  \\
  HD  60498& 247.20&  -6.64&     936&       1&        --  &         -- &        --  &         -- &       -41.2&       -41.2&       IVC  \\
  HD  60498& 247.20&  -6.64&     936&       2&        --  &         -- &        --  &         -- &       -31.2&       -31.2&       IVC  \\
  HD  60498& 247.20&  -6.64&     936&      12&        --  &         -- &        --  &         -- &        53.4&        44.3&       IVC  \\
  HD  68761& 254.37&  -1.62&    1473&       1&        --  &         -- &        --  &         -- &       -55.3&       -55.3&       IVC  \\
  HD  68761& 254.37&  -1.62&    1473&       2&        --  &         -- &        --  &         -- &       -39.2&       -39.2&       IVC  \\
  HD  68761& 254.37&  -1.62&    1473&      10&        --  &         -- &        --  &         -- &        50.8&        38.8&       IVC  \\
  HD  74966& 258.08&   3.93&     658&       7&        --  &         -- &        --  &         -- &        35.7&        31.8&       IVC  \\
  HD  74966& 258.08&   3.93&     658&       8&        --  &         -- &        --  &         -- &        50.5&        46.6&       IVC  \\
  HD  72067& 262.08&  -3.08&     488&       1&        --  &         -- &        --  &         -- &      -105.8&      -105.8&       HVC  \\
  HD  89587& 279.83&   5.19&     900&       1&        --  &         -- &        --  &         -- &       -40.1&       -37.4&       IVC  \\
  HD  93131& 287.67&  -1.08&    2600&       1&        --  &         -- &        --  &         -- &       -67.4&       -57.1&       IVC  \\
  HD  93131& 287.67&  -1.08&    2600&       2&        --  &         -- &        --  &         -- &       -51.4&       -41.0&       IVC  \\
  HD  94910& 289.18&  -0.69&    6000&       1&      -117.3&      -105.1&        --  &         -- &        --  &         -- &       HVC  \\
  HD  94910& 289.18&  -0.69&    6000&       2&       -79.4&       -67.2&        --  &         -- &        --  &         -- &       IVC  \\
  HD  94910& 289.18&  -0.69&    6000&       3&       -64.5&       -52.3&       -66.2&       -54.0&        --  &         -- &       IVC  \\
  HD  96917& 289.28&   3.06&    2910&       1&        --  &         -- &        --  &         -- &       -70.8&       -58.5&       IVC  \\
  HD  96917& 289.28&   3.06&    2910&      11&        --  &         -- &        --  &         -- &        60.1&        60.1&       IVC  \\
  HD  94963& 289.76&  -1.81&    3257&       1&        --  &         -- &        --  &         -- &       -89.0&       -76.1&       IVC  \\
  HD  94963& 289.76&  -1.81&    3257&       2&        --  &         -- &        --  &         -- &       -54.0&       -41.0&       IVC  \\
  HD  94963& 289.76&  -1.81&    3257&       3&        --  &         -- &        --  &         -- &       -44.8&       -31.9&       IVC  \\
  HD  97253& 290.79&   0.09&    2415&       1&        --  &         -- &        --  &         -- &       -47.4&       -33.7&       IVC  \\
  HD  29138& 297.99& -30.54&    3530&       1&        --  &         -- &        --  &         -- &       -77.6&       -56.9&       IVC  \\
  HD 106068& 298.51&  -0.41&    2824&       1&        --  &         -- &        --  &         -- &       -85.0&       -61.3&       IVC  \\
  HD 106068& 298.51&  -0.41&    2824&       2&        --  &         -- &        --  &         -- &       -75.6&       -51.9&       IVC  \\
  HD 109867& 301.71&  -4.35&    3264&       1&        --  &         -- &       -63.0&       -33.3&       -63.1&       -33.4&       IVC  \\
  HD 136239& 321.23&  -1.75&    2350&       1&        --  &         -- &       -65.2&       -31.6&       -64.2&       -30.6&       IVC  \\
  HD 148937& 336.37&  -0.22&    1380&       1&        --  &         -- &        --  &         -- &      -102.4&       -87.3&       IVC  \\
  HD 148688& 340.72&   4.35&    1452&       1&        --  &         -- &       -41.7&       -28.0&       -46.1&       -32.4&       IVC  \\
  HD 148688& 340.72&   4.35&    1452&       8&        --  &         -- &        --  &         -- &        34.1&        34.1&       IVC  \\
  HD 151932& 343.22&   1.43&    1413&       1&        --  &         -- &        --  &         -- &       -56.2&       -44.4&       IVC  \\
  HD 151932& 343.22&   1.43&    1413&       2&        --  &         -- &        --  &         -- &       -44.0&       -32.2&       IVC  \\
  HD 151932& 343.22&   1.43&    1413&      11&        --  &         -- &        --  &         -- &        51.5&        51.5&       IVC  \\
  HD 152235& 343.31&   1.10&    1861&       1&        --  &         -- &        --  &         -- &       -75.3&       -58.8&       IVC  \\
  HD 152235& 343.31&   1.10&    1861&      13&        --  &         -- &        --  &         -- &        39.8&        39.8&       IVC  \\
  HD 152003& 343.33&   1.41&    2687&       1&        --  &         -- &        --  &         -- &       -62.9&       -36.3&       IVC  \\
  HD 152270& 343.49&   1.16&    1900&      10&        --  &         -- &        --  &         -- &        31.7&        31.7&       IVC  \\
  HD 157038& 349.95&  -0.79&    1444&       1&        --  &         -- &        --  &         -- &       -87.3&       -79.6&       IVC  \\
  HD 157038& 349.95&  -0.79&    1444&       2&        --  &         -- &        --  &         -- &       -69.6&       -61.9&       IVC  \\
  HD 157038& 349.95&  -0.79&    1444&       3&        --  &         -- &        --  &         -- &       -59.1&       -51.4&       IVC  \\
  HD 157038& 349.95&  -0.79&    1444&       4&        --  &         -- &       -47.8&       -40.1&       -48.9&       -41.2&       IVC  \\
  HD 155806& 352.59&   2.87&    1064&       1&        --  &         -- &        --  &         -- &       -37.0&       -33.0&       IVC  \\
  HD 163758& 355.36&  -6.10&    4103&       1&        --  &         -- &        --  &         -- &       -46.9&       -30.8&       IVC  \\
  HD 163758& 355.36&  -6.10&    4103&      10&        --  &         -- &        --  &         -- &        38.9&        38.9&       IVC  \\
  HD 163758& 355.36&  -6.10&    4103&      11&        --  &         -- &        --  &         -- &        90.4&        90.4&       HVC  \\
           &       &       &        &        &            &            &            &            &            &            &            \\
\hline
\end{tabular}
\normalsize
\end{center}
\end{table*}

\section{Conclusions}                                      \label{s_conclusions}

We have obtained spectra towards 74 early-type stellar targets from the
UVES POP survey, at a velocity resolution of 3.75\,km~s$^{-1}$. We
have studied the interstellar Na\,{\sc i} UV doublet at 3302\,\AA, 
Ti\,{\sc ii} at 3383\,\AA\ and Ca\,{\sc ii}
at 3933\,\AA. The average S/N ratio per pixel for the spectra are 260, 300 and 430,
respectively, which far exceeds that in previous work in a sample of this size. 
Column densities were measured for each component along each line-of-sight,
 with the individual components being summed to obtain the total. This quantity
was also
calculated by the AOD method, which is in excellent agreement with the
total generated by the fitting of each individual component. There is also
good
agreement between the total column densities derived from each line of the Na\,{\sc i} UV 
doublet.  The median of the errors in individual components are 
0.06~dex, 0.06~dex and 0.02~dex for Na, Ti and Ca components respectively.
In general, we have not studied cases where the stellar
line is of a similar width and velocity to the interstellar profile, and so we have avoided
stellar contamination of our derived column densities.

\subsection{General correlations}                             \label{s_con_corr}

We have investigated the
dependence of the total column density of each species along the line-of-sight
to a star with distance, reddening and total H\,{\sc i} column density. 
Our stars span distances from 52\,pc to 6000\,pc, with the
maximum height from to the Galactic plane, $|$z$|$, being 1793\,pc. We find
a strong positive correlation between Ti and Ca column densities
and distance, and a similar correlation for Na although with much more scatter.
We also find a good correlation between the column densities and 
$|z|$, although again the correlation is weaker in Na. Our sightlines have
values of reddening $E(B-V)$ of up to 1.0, and some correlation between
the column density per parsec of each species and the reddening per kiloparsec has been found. We find similar correlations between the H\,{\sc i} column
density per parsec and the column density per parsec of Ca and Ti. There is evidence of a stronger correlation between Na and H\,{\sc i} but
given the scatter of the points this may not be significant. Wakker \& Mathis (2000) have previously 
found this large dispersion, and have thus cautioned against for example using non-detections of Na 
to set limits on the distances of High Velocity Clouds. The observed scatter in Na and H\,{\sc i} is 
attributed by these authors to the fact that it is not the dominant ionisation stage in the ISM, 
so ionisation effects may play a more important role. This certainly
calls into question the accuracy of using the Na column density as an
indicator of the value for H\,{\sc i}. We also find that the column
density derived from the Na\,{\sc i} D lines, generally used to predict the H\,{\sc i}
values, appears to underestimate the column density relative to the
UV lines that we have studied. Even when the saturation corrections of
Savage \& Sembach \cite{sav91} are applied to the column density derived from the Na\,{\sc i} D lines it still appears
to underestimate the Na column density. For Na column densities greater than 13\,dex the corrected values
underestimate the column density by as much as 0.3\,dex relative to column density derived from the UV lines.

\subsection{Depletions}                                        \label{s_con_dep}

The depletion of each species relative to the solar system
value has been correlated with $|z|$ and $E(B-V)$. We find some 
evidence for decreasing depletion with increasing $z$ for both Ti and Ca,
with no correlation with depletion and reddening for these species. For Na, 
we find decreasing depletion with increasing E(B--V), some of which may 
be caused by the conversion of H\,{\sc i} to H$_{2}$.

\subsection{$N$(Ti\,{\sc ii})/$N$(Ca\,{\sc ii}) ratio}           \label{s_con_Ti/Ca}

We find a good correlation between the Ti\,{\sc ii} and Ca\,{\sc ii} column
densities, both in the total values and those from
individual components that are observed in both species. The ratio from the
total is 0.30 compared to that from the individual components of 0.26. Our
ratios compare very well
with the value of 0.29~$\pm$~0.03
calculated by W97. The similarity between our plots of the
correlation of Ti and Ca 
with each of the different parameters
reinforces the possibility that the species show good correspondence 
if the Ca ionisation balance is controlled by the average interstellar 
radiation field, and the relative Ca and Ti depletions are relatively 
constant in the clouds (Welty et al. 1996, Wakker \& Matthis 2000). 

\subsection{Future work}

As discussed in Section 4.4.1, many of our sightlines show strong IVC and 
HVC features. A future paper will involve a more detailed analysis of these 
components, using the Villa-Elissa H\,{\sc i} survey and the current 
observations to provide improved distance limits to IVCs and HVCs. 

\section*{acknowledgements}                            \label{s_acknowledgments}

We would like to thank the staff of the Very Large Telescope, Paranal for 
the work involved in producing the POP survey.
ESO DDT programme ID 266.D-5655, http://www.eso.org/uvespop. IH and JVS gratefully 
acknowledge the ESO Director Generals Discretionary fund which provided financial 
support for IH for this work. FPK is grateful to AWE Aldermaston for the
award of a William Penney Fellowship.
This research has made use of the {\sc simbad} database, 
operated at CDS, Strasbourg, France. We would also like to thank the referee who provided 
many useful comments and suggestions to the initial draft of the paper.


{}


\begin{thebibliography}{}

\bibitem[1993]{alb93}Albert C.~E., Blades J.~C., Morton D.~C., Lockman F.~J., Proulx M., Ferrarese L.,
1993,  ApJS, 88, 81

\bibitem[1995]{bal95}Ballereau D., Chauville J., Zorec J., 1995, A\&AS, 111, 
423

\bibitem[2003]{bag03}Bagnulo S., Jehin E., Ledoux C., Cabanac R., Melo C.,
Gilmozzi R., 2003, ESO Messenger no. 114 Page 10

\bibitem[1999]{bra99} Braun R., Burton W.~B., 1999, 
A\&A, 341, 437

\bibitem[2000]{cal00}Callaway M.~B., Savage B.~D., Benjamin R.~A., Haffner L.~M., Tufte S.~L., 2000, ApJ, 532, 943

\bibitem[1999]{cha99}Cha A.~N., Sembach K.~R., Danks A.~C., 1999, ApJ, 515, 25

\bibitem[1983]{con83}Conti P.~S., Garmany C.~D., de Lore C., Vanbeveren D.,
1983, ApJ, 274, 302

\bibitem[1990]{con90}Conti P.~S., Vacca W.~D., 1990, AJ, 100, 431

\bibitem[1994]{cri94} Crinklaw G., Federman S.~R., Joseph C.~L., 1994, 
ApJ, 424, 748

\bibitem[1994]{dip94} Diplas A., Savage B.~D., 1994, ApJS, 93, 211

\bibitem[1989]{edg89}Edgar R.~J. Savage B.~D., 1989, ApJ, 340, 762

\bibitem[1985]{fer85}Ferlet R., Vidal-Madjar A., Gry C., 1985, ApJ, 298, 838

\bibitem[1998]{fri98} Frisch P.~C., 1998,
LNP, 506, 269

\bibitem[1983]{fri83} Frisch P.~C., York D.~G., 1983,
ApJ, 271, 59

\bibitem[1994]{fru94}Fruscione A., Hawkins I., Jelinsky P., Wiercigroch A., 
1994, ApJS, 94, 127

\bibitem[1985]{gon85} Gondhalekar P.M., 1985, ApJ, 293, 230

\bibitem[1988]{gor88}Goraya P.~S., Tur N.~S., 1988, AJ, 96, 346

\bibitem[1987]{gra87}Grady C~.A., Bjorkman K.~S., Snow T.~P., 1987, ApJ, 320,
 376

\bibitem[1904]{har04} Hartmann J., 1904,
ApJ, 19, 268

\bibitem[1974]{hob74}Hobbs L.~M., 1974, ApJS, 191, 381

\bibitem[1978]{hob78}Hobbs L.~M., 1978, ApJS, 38, 129

\bibitem[1984]{hob84}Hobbs L.~M., 1984, ApJS, 56, 315

\bibitem[1993]{hoe93}Hoekzema N.~M., Lamers H.~J.~G.~L.~M, van Genderen A.~M., 
1993, A\&AS, 98, 505

\bibitem[2003]{how03} Howarth I.~D., Murray J., Mills D., Berry D.~S., 2003,
      Starlink User Note SUN 50, Rutherford Appleton Laboratory/CCLRC
      
\bibitem[1996]{hur96}Hurwitz M., Bowyer S., 1996, ApJ, 465, 296

\bibitem[1997]{kre97} Krelowski J., Schmidt M., Snow T.~P., 1997, PASP, 109, 1135

\bibitem[1987]{lei87} Leitherer C., Chavarria-K C., 1987, A\&A, 175, 208

\bibitem[2003]{lal03} Lallement R., Welsh B.~Y., Vergely L., Crifo F., Sfeir D., 2003,
A\&A, 411, 447

\bibitem[2003]{lod03} Lodders K., 2003, ApJ, 591, 1220

\bibitem[1994]{men94} Mennickent R.~E., Vogt N., Barrera L.~H., Covarrubias
 R., Ramirez A., 1994, A\&AS, 106, 427

\bibitem[1981]{mih81} Mihalas D., Binney J., 1981, {\em Galactic Astronomy;
Structure and Kinematics}, 2nd edition, San Francisco, W. H. Freeman and Company

\bibitem[2000]{mor00} Morras R., Bajaja E., Arnal E.~M., Poppel W.~G.~L., 2000,
A\&AS, 142, 25

\bibitem[2003]{mor03} Morton D.~C., 2003, ApJS, 149, 205

\bibitem[2004]{mor04} Morton D.~C., 2004, ApJS, 151, 403

\bibitem[1996]{pen96} Penny L.~R., 1996, ApJ, 463, 737

\bibitem[1984]{phi84} Phillips A.P., Pettini M., Gondhalekar P.M., 1984,
MNRAS, 206, 337

\bibitem[1990]{sav90} Savage B.~D., Edgar R.~J., Diplas A., 1990, ApJ, 361,   
107

\bibitem[1991]{sav91} Savage B.~D., Sembach K.~R., 1991, ApJ, 379, 245

\bibitem[1982]{sch82} Schmidt-Kaler Th., 1982, in Schaifers K., Voigt H.~H., 
eds, Landolt-Boernstein, group VI, subvol. b, Stars and Star Clusters, Vol. 2
Springer, Berlin, p. 17

\bibitem[1993]{sem93} Sembach K.~R., Danks A.~C., Savage B.~D., 1993, 
A\&AS, 100, 107 (SDS93)

\bibitem[1994]{sem94} Sembach K.~R., Danks A.~C., 1994,
A\&A, 289, 539 (SD94)

\bibitem[2000]{sem00} Sembach K.~R., Howk J.~C., Ryans R.~S.~I., Keenan F.~P., 2000,
ApJ, 528, 310

\bibitem[2004]{sla04} Slavin J.~D., Nichols J.~S., Blair W.~P., 2004, 
ApJ, 606, 900

\bibitem[1982]{sle82} Slettebak A., 1982, ApJS, 50, 55

\bibitem[2003]{smo03} Smoker J.~V., et al., 2003, 
MNRAS, 346, 119

\bibitem[1985]{spi85} Spitzer L.~J., 1985, 
ApJ, 290, 21

\bibitem[1978]{sto78} Stokes G.~M., 1978, ApJS, 36, 115

\bibitem[1996]{vac96} Vacca W.~D., Garmany C.~D., Shull J.~M., 1996,
ApJ, 460, 914

\bibitem[1993]{val93} Vallerga J.~V., Vedder P.~W., Craig N., Welsh B.~Y., 1993,
ApJ, 411, 729

\bibitem[2000]{ver00} Vergely J.~-L., Freire Ferro R., Siebert A., Valette B.,
2000, A\&A 366, 1016

\bibitem[1991]{wak91}Wakker B.~P., 1991, A\&A, 250, 499

\bibitem[1997]{wak97}Wakker B.~P., van Woerden H., 1997,
ARA\&A, 35, 217

\bibitem[2000]{wak00} Wakker B.~P., Mathis J.~S., 2000, 
ApJ, 544, 107

\bibitem[1996]{wal96} Wallace P., Clayton C., 1996, {\sc rv}, Starlink User Note
SUN 78, Rutherford Appleton Laboratory/CCLRC

\bibitem[1994]{weg94} Wegner, W., 1994, MNRAS, 270, 229

\bibitem[2000]{weg00}Wegner, W., 2000, MNRAS, 319, 771

\bibitem[1997]{wel97} Welsh B.~Y., Sassen T., Craig N., Jelinsky S., Albert
C.~E., 1997, ApJS, 112, 507 (W97)

\bibitem[1994]{welt94} Welty D.~E., Hobbs L.~M., Kulkarni V.~P., 1994,
ApJ, 436, 152

\bibitem[1996]{wel96} Welty D.~E., Morton D.~C., Hobbs L.~M., 1996, 
ApJS, 106, 533 (WMH96)

\bibitem[1997]{win97} Winkler H., 1997, MNRAS, 287, 481

\end{thebibliography}
\end{document}